\documentclass[a4paper]{article}
\pdfoutput=1 

\usepackage{jheppub} 

\usepackage[T1]{fontenc} 
\usepackage[utf8]{inputenc} 
\usepackage{microtype} 
\usepackage{lmodern} 
\usepackage{booktabs} 
\usepackage{mdframed} 
\usepackage{graphicx}
\usepackage{todonotes} 
\usepackage{slashed}
\usepackage{color}
\usepackage{cancel}
\usepackage[shortlabels]{enumitem}
\usepackage{tikz}
\usepackage{young}
\usepackage[vcentermath]{youngtab}
\usepackage{tensor}
\usepackage[normalem]{ulem} 

\providecommand*{\dd}{\mathop{}\!d}
\renewcommand*{\dd}{\mathop{}\!d}

\providecommand*{\pd}{\mathop{}\!\partial}
\renewcommand*{\pd}{\mathop{}\!\partial}

\providecommand*{\hs}{\mathop{}\!*}
\renewcommand*{\hs}{\mathop{}\!*}

\providecommand*{\curl}{\mathop{}\!\mathrm{curl}}
\renewcommand*{\curl}{\mathop{}\!\mathrm{curl}}

\providecommand*{\R}{{\mathbb{R}}}
\renewcommand*{\R}{{\mathbb{R}}}

\newcommand{\cB}{\mathcal{B}} 
\newcommand{\mcB}{\CMcal{B}} 
\newcommand{\et}{{\widetilde{\epsilon}}}
\newcommand{\cN}{\CMcal{N}}

\newcommand{\tta}{\mathtt{a}}
\newcommand{\ttb}{\mathtt{b}}
\newcommand{\ttc}{\mathtt{c}}
\newcommand{\ttd}{\mathtt{d}}
\newcommand{\tte}{\mathtt{e}}

\newcommand{\bbar}[1]{\bar{\bar{ #1 }}}

\newcommand{\onetab}[1]{\begin{tabular}{c} #1 \end{tabular}}

\usepackage{pdflscape}
\usepackage{multirow}
\newcommand{\otoprule}{\midrule[\heavyrulewidth]}

\usepackage[smalltableaux,centertableaux,centerboxes]{ytableau}

\usepackage{hyperref}
 \usepackage{color}
 \definecolor{dark-red}{rgb}{0.4,0.15,0.15}
 \definecolor{dark-blue}{rgb}{0.15,0.15,0.4}
 \definecolor{medium-blue}{rgb}{0,0,0.5}
 \hypersetup{colorlinks, linkcolor={dark-red}, citecolor={dark-blue}, urlcolor={medium-blue}}

\usepackage{amsmath,amsfonts,amssymb}
\usepackage{eucal} 
\usepackage{mleftright} \mleftright
\usepackage{tensor} 
\usepackage{braket} 
\usepackage{mathrsfs} 

\usepackage{cleveref}
\crefname{equation}{}{}

\providecommand*{\dd}{\mathop{}\!d}
\renewcommand*{\dd}{\mathop{}\!d}

\providecommand*{\pd}{\mathop{}\!\partial}
\renewcommand*{\pd}{\mathop{}\!\partial}

\providecommand*{\hs}{\mathop{}\!*}
\renewcommand*{\hs}{\mathop{}\!*}

\providecommand*{\curl}{\mathop{}\!\mathrm{curl}}
\renewcommand*{\curl}{\mathop{}\!\mathrm{curl}}

\providecommand*{\R}{{\mathbb{R}}}
\renewcommand*{\R}{{\mathbb{R}}}

\title{\boldmath The Action of the (Free) $\cN = (3,1)$ Theory in Six Spacetime Dimensions}

\author[a,b,1]{Marc Henneaux\note{\href{https://orcid.org/0000-0002-8912-6384}{ORCID: 0000-0002-8912-6384}},}
\author[a,2]{Victor Lekeu\note{\href{https://orcid.org/0000-0001-6111-005X}{ORCID: 0000-0001-6111-005X}},}
\author[a,3]{Javier Matulich\note{\href{ https://orcid.org/0000-0002-3558-9025}{ORCID: 0000-0002-3558-9025}},}
\author[a,4]{and Stefan Prohazka\note{\href{https://orcid.org/0000-0002-3925-3983}{ORCID: 0000-0002-3925-3983}}}

\affiliation[a]{Universit\'e Libre de Bruxelles and International Solvay Institutes, 
  Campus Plaine---CP~231, \\
  B-1050  Bruxelles, Belgium}
\affiliation[b]{Coll\`ege de France, 11 place Marcelin Berthelot, F-75005 Paris, France}

\emailAdd{henneaux@ulb.ac.be}
\emailAdd{vlekeu@ulb.ac.be}
\emailAdd{jmatulic@ulb.ac.be}
\emailAdd{stefan.prohazka@ulb.ac.be}

\abstract{
The action of the free $\cN = (3,1)$ theory in six spacetime dimensions is explicitly constructed. The variables of the variational principle are prepotentials adapted to the self-duality conditions on the fields.  The $(3,1)$ supersymmetry variations are given and the invariance of the action is verified.  The action is first-order in time derivatives.  It is also Poincar\'e invariant but not manifestly so, just like the Hamiltonian action of  more familiar relativistic field theories. 
}

\begin{document} 
\maketitle
\flushbottom

\newpage

\section{Introduction}
\label{sec:introduction}

The maximal supersymmetry algebra is unique in all spacetime dimensions $4 \leq D \leq 11$, except in dimensions $6$ and $10$, where one can independently assign different chiralities to the supercharges
\cite{Nahm:1977tg,Strathdee:1986jr}.  These spacetime dimensions are also the dimensions where chiral $\left(\frac{D}{2}-1\right)$-forms can be consistently defined\footnote{We restrict the spacetime dimension $D$ to $4 \leq D \leq 11$ because we are interested here in the various higher dimensional origins of $D = 4$, $\cN = 8$ supergravity.  Dimension $2$ is of course also very special.}.

\begin{figure}
\centering
\begin{tikzpicture}
\tikzset{r/.style={rectangle, rounded corners, text centered, draw}}
\tikzset{rt/.style={rectangle, rounded corners, text centered, draw, ultra thick}}
\node[r] (11) at (0,14) {$D = 11$ supergravity};
\node[r] (A) at (0,12) {\onetab{$D=10$, type $II_A$\\non-chiral $\cN = (1,1)$\\$G=\R$}};
\node[r] (9) at (0,10) {\onetab{$D=9$\\$G = SL(2,\R) \times \R$}};
\node[r] (6) at (0,6) {\onetab{$D=6$\\non-chiral $\cN = (2,2)$\\$G = SO(5,5)$}};
\node[r] (5) at (0,4) {\onetab{$D=5$\\$G=E_{6(6)}$}};
\node[r] (4) at (0,2) {\onetab{$D=4$\\$G=E_{7(7)}$}};
\node[r] (B) at (-4.5,12) {\onetab{$D=10$, type $II_B$\\ chiral $\cN = (2,0)$\\$G=SL(2,\R)$}};
\node[r] (40) at (-4.5,6) {\onetab{$D=6$ exotic theory\\chiral $\cN = (4,0)$\\$G=E_{6(6)}$}};
\node[rt] (31) at (4.5,6) {\onetab{$D=6$ exotic theory\\ chiral $\cN = (3,1)$\\$G=F_{4(4)}$}};
\foreach \from/\to in {11/A, A/9, B/9, 6/5, 5/4}
	\draw [->, shorten >=3pt] (\from) -- (\to);
\foreach \from/\to in {40/5, 31/5}
	\draw [->, dashed, shorten >=3pt] (\from) -- (\to);
\draw (9) -- (0,9);
\draw[dashed] (0,9) -- (0,7.5);
\draw[->, shorten >=3pt] (0,7.5) -- (6);
\end{tikzpicture}
\caption{\label{fig:oxidation}The various higher-dimensional origins of $D = 4$, $\cN = 8$ supergravity, with the $\cN = (3,1)$ theory highlighted. We have indicated the supersymmetry in dimensions ten and six, where chirality allows for different maximal supersymmetry algebras. The global symmetry groups $G$ (conjectured in \cite{Hull:2000zn} in the case of six-dimensional exotic theories) are also written.}
\end{figure}
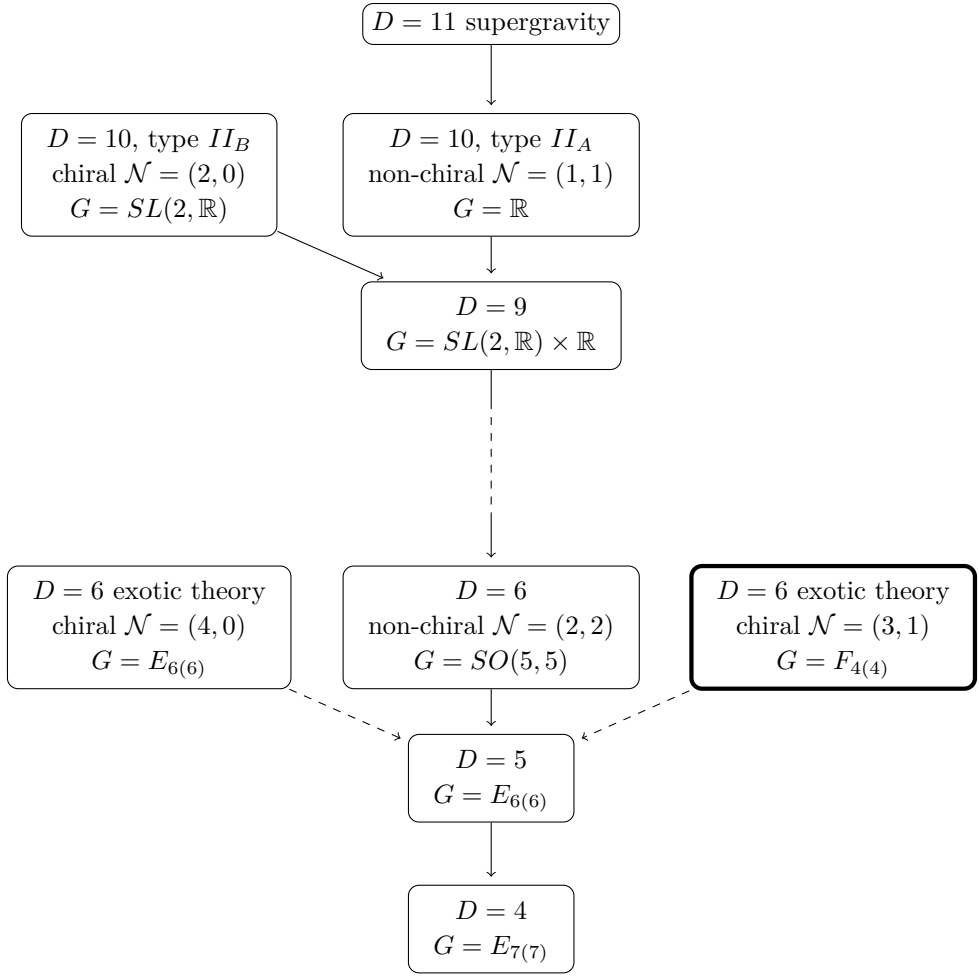

In $D=10$ spacetime dimensions, there are two maximal supersymmetry algebras, the chiral $\cN = (2,0)$ algebra and the non-chiral $\cN = (1,1)$ algebra.  In $\cN = (\cN_+, \cN_-)$,  the number $\cN_+$ (respectively, $\cN_-)$ denotes the number of supercharges of positive (respectively, negative) chirality.  The theory realizing the non-chiral supersymmetry algebra is type $II_A$ supergravity, which is the Kaluza-Klein reduction of maximal supergravity in $11$ dimensions.  The theory realizing the chiral supersymmetry algebra is type $II_B$ supergravity. Although involving a chiral $4$-form\footnote{Due to the self-duality condition satisfied by the corresponding field strength, the construction of an action
principle is notoriously subtle~\cite{Marcus:1982yu}.}, one can formulate an action principle, which is covariant but not manifestly so, by handling the chiral $4$-form along the methods of \cite{Henneaux:1988gg} (see \cite{Bekaert:1999sq} for the explicit derivation; see also \cite{DallAgata:1997gnw,DallAgata:1998ahf} for a covariant PST-like formulation \cite{Pasti:1995ii,Pasti:1995tn,Pasti:1996vs} involving extra fields and non-polynomial interactions). Accordingly, both algebras are realized by non trivial interacting theories.  These theories reduce to the same, unique maximal supergravity in $D=9$ spacetime dimensions (see Figure \ref{fig:oxidation}).

The situation is more intricate in $D=6$ spacetime dimensions, where three different maximal supersymmetry algebras exist: the $(4,0)$ and $(3,1)$ chiral algebras and the $(2,2)$ non-chiral algebra.  While the theory realizing the $(2,2)$ supersymmetry algebra is well known and just the toroidal dimensional reduction of maximal supergravity in $11$ dimensions, the theories realizing the other two superalgebras (if they exist) are more mysterious.  This is because they would involve, in place of the standard spin-2 field describing gravity, tensor fields with mixed Young symmetries subject to chirality conditions. Hence the name ``exotic (super)gravity''.  In view of the subtleties for writing an action principle for chiral bosonic fields, and the various no-go theorems preventing the interactions of tensor fields with mixed Young symmetries \cite{Bekaert:2002uh,Boulanger:2004rx,Bekaert:2004dz,Ciobirca:2004yu,Bizdadea:2009zc} or involving chirality conditions \cite{Bekaert:1999dp,Bekaert:1999sq,Bekaert:2000qx}, the maximal chiral supersymmetry algebras were largely ignored.

A notable exception is the work  \cite{Hull:2000zn,Hull:2000rr,Hull:2001iu}, in which it was argued that the $(4,0)$ theory could emerge as the strong coupling limit of theories having maximal supergravity as their low energy effective theory in five spacetime dimensions.  Intriguing (and remarkable) properties of this putative theory were also indicated, and  the free equations of motion of the various fields occurring in the $(4,0)$ multiplet were given.  Similar intriguing features of the other putative $(3,1)$ theory were discussed in that same reference, as well as the corresponding free equations of motion. 

The conjectured exotic $(4,0)$ supergravity was further argued in \cite{Anastasiou:2013hba,Borsten:2017jpt,Anastasiou:2017taf} to find a natural place in a ``conformal magic pyramid'', and to be obtained by squaring two maximal tensor multiplets of same chirality.

Now,  in order to define a theory, one needs more than just the equations of motion.  One also needs the commutation relations between the dynamical variables, i.e., in the classical limit, their Poisson brackets. The information on both the equations of motion and the (pre-)symplectic  structure, which encodes the Poisson bracket structure, is contained in the action.  It is therefore important to write the action when it exists.

In a recent paper \cite{Henneaux:2017xsb}, the description of the free $(4,0)$ theory was completed by showing that an action principle did exist through an explicit construction.  The supersymmetry transformations of the fields were also given and the invariance of the action was established.  The action involves prepotentials, as in \cite{Henneaux:2004jw}.  These prepotentials are adapted to the duality properties of the theory but they are spatial objects.  For that reason, although present, spacetime covariance is not manifest.   The situation is identical to the Hamiltonian formulation of familiar relativistic field theories and holds also, as we recalled, for the variational formulation of type $II_B$ supergravity.

In this paper, we focus on the free $(3,1)$ theory.  We derive an explicit action principle, which also involves prepotentials and is again ``intrinsically Hamiltonian'': it is of first order in the time derivatives and covariant although not manifestly so.  We  write down the supersymmetry transformations and verify the invariance of the action.   The construction of the action  completes the definition of the free theory and  fills at the linear level the remaining box in the chain of higher dimensional parents of $D = 4$, $\cN = 8$ supergravity of Figure \ref{fig:oxidation}.

Our paper is organized as follows.  The multiplet of the $D = 6$, $\cN = (3,1)$ theory contains
a chiral gauge field $\phi_{\mu\nu\rho}$ with $(2,1)$ Young symmetry~\cite{Hull:2000zn}.  The action for  a non-chiral $(2,1)$ Young symmetry field was written in \cite{Curtright:1980yk}. The implementation of the self-duality condition satisfied by the corresponding curvature is more intricate than for ordinary $p$-forms.  Section \ref{sec:Chiral21} explains how to overcome 
the obstacle.   First, we prove that the self-duality condition is equivalent to a set of equations containing only the spatial components $\phi_{ijk}$: schematically,
\begin{equation}
  R = \hs R
  \quad\Leftrightarrow\quad
  \mathcal{E} = \mathcal{B}
  \quad\Leftrightarrow\quad
  \curl(\mathcal{E} - \mathcal{B})=0 \, \text{ and } \, \bar{\bar{{\mathcal E}}}=0 \, ,
\end{equation}
where $\mathcal{E}$ and $\mathcal{B}$ denote the electric and magnetic fields of $\phi_{\mu\nu\rho}$, respectively, and where the bar stands for the trace (see section \ref{sec:electr-magn-field} for details).
In addition to the dynamical equation, the last set contains a constraint on the $\phi_{ijk}$ variables (double tracelessness of the electric field).  Second, we show that this constraint can be solved by expressing the field in terms of a new variable $Z$, called the prepotential. In terms of $Z$, the equation $\curl(\mathcal{E} - \mathcal{B})=0$ becomes
\begin{equation}
\dot D[Z] = \curl \, D[Z]\, ,
\end{equation}
where $D[Z]$ is the Cotton tensor of $Z$ defined in section \ref{sec:chiralaction}.
 These final equations have the advantage that they allow for a 
remarkably simple variation principle, given by\footnote{We ignore possible boundary contributions in this work.}
\begin{equation}
  S[Z] = \frac{1}{2}\int  \dd t\dd^5 \! x \,
  Z_{ijkpq}\left(\dot{D}^{ijkpq}[Z] -
      \frac{1}{2}
      \varepsilon^{abcpq}\pd_{a}D\indices{^{ijk}_{bc}}\left[Z\right]
    \right)
    \,.
    \label{actionZintro}
\end{equation}
This whole procedure closely follows the construction of variational
principles for chiral $p$-forms~\cite{Henneaux:1988gg} and chiral gauge
fields with $(2,2)$ Young symmetry~\cite{Henneaux:2016opm}. It has the characteristic form of the actions where duality plays a crucial role \cite{Henneaux:2016zlu,Henneaux:2016opm,Lekeu:2018kul}.

Although not manifestly so, this action is Poincaré invariant and leads to equations of motion physically equivalent to the original (Poincaré invariant) equations $R = \hs R$. This is a usual feature of this kind of first-order actions and was already discussed in \cite{Henneaux:1988gg} (see also \cite{Bunster:2012hm} for the complementarity between duality and Poincaré invariance). It could be interesting to see if one could add (non-quadratic) auxiliary fields to this action to restore manifest Poincaré invariance along the lines of \cite{Pasti:1995ii,Pasti:1995tn,Pasti:1996vs}, but we will not do so here.

Once the variational principle for the exotic chiral $(2,1)$ Young symmetry field is understood, the construction of the complete action of the free $\cN = (3,1)$ theory is straightforward. This is achieved in Section \ref{sec:Action}.  We then give in Section \ref{sec:susy} the supersymmetry variations of the prepotentials and verify the invariance of the action.   Section \ref{sec:Conclu} is devoted to final comments and conclusions.

Various appendices complete the analysis.  Appendix \ref{app:A}  provides the technical steps necessary to prove the equivalence of the covariant self-duality equations on the curvature with various forms of the equations expressed in terms of electric and magnetic fields.  Appendix \ref{sec:conf} develops the necessary algebraic tools for handling  the conformal geometry of the prepotentials.  Appendix \ref{app:ham} is devoted to an alternative derivation of the  action, which can be obtained by splitting the action for a non-chiral $(2,1)$ tensor into its chiral and anti-chiral parts, following \cite{Deser:1997se,Bekaert:1998yp}. Appendices \ref{sec:dimred} and \ref{app:susy} establish the connection with (linearized) maximal five-dimensional supergravity, which is shown to be correctly reproduced upon dimensional reduction to five dimensions.

\paragraph{Conventions.}

The spacetime is flat Minkowski space $\R^{5,1}$
with a Lorentzian metric of ``mostly plus signature''.
The epsilon tensor $\varepsilon$ is totally antisymmetric and such that $\varepsilon_{012345}=+1$. For the space/time split and dimensional reduction, this implies $\varepsilon^{(5)}_{ijklm}
=\varepsilon^{(5,1)}_{0ijklm}$, $\varepsilon^{(4,1)}_{\mu\nu\rho\sigma\tau}
=\varepsilon^{(5,1)}_{\mu\nu\rho\sigma\tau5}$ and $\varepsilon^{(4)}_{ijkl} = \varepsilon^{(5,1)}_{0ijkl5}$ (the superscripts denote the spacetime signature, but we will not write them explicitly).

A tensor with $(a_{1},a_{2},\cdots,a_{n})$ Young symmetry
is labeled by the length of the rows, i.e., corresponds to a Young diagram with $n$ rows which have $a_{i}$ boxes.
We follow the manifestly antisymmetric convention.

For gamma matrices and (symplectic Majorana-Weyl) spinors, we follow the conventions and notations of \cite{Henneaux:2017xsb} (see also \cite{Lekeu:2018kul} for useful identitites in five dimensions). The same is true for the $\mathfrak{usp}(2N)$ algebras. 

\section{The chiral \texorpdfstring{$(2,1)$}{(2,1)}-tensor}
\label{sec:Chiral21}

\subsection{Equations of motion}

We start by reviewing the chiral gauge field $\phi_{\alpha_1 \alpha_2 \beta_1}$ of $(2,1)$ Young symmetry type in $D=5+1$ spacetime dimensions,  considered by Hull in \cite{Hull:2000zn}. This field reduces in $5$ spacetime dimensions to the standard Pauli-Fierz field plus a standard massless vector field (see Appendix \ref{sec:dimred}).  It provides for that reason what can be regarded as an exotic description of the combined Einstein-Maxwell system.  This exotic description is different from that given by the chiral gauge field of the $\cN = (4,0)$ theory, which has $(2,2)$ Young symmetry type, and which contains only an ``exotic graviton''.   By constrast, the Pauli-Fierz field $h_{\mu \nu} = h_{\nu \mu}$ in $6$ spacetime dimensions gives one graviton, one photon {\em and} one massless scalar upon Kaluza-Klein reduction to $5$ spacetime dimensions.

In terms of Young tableaux we will denote the exotic graviton-photon $\phi_{\alpha_1 \alpha_2 \beta_1}$ by
\begin{equation}
    \phi_{\alpha_1 \alpha_2 \beta_1} 
\sim 
\begin{ytableau}
  \alpha_1 & \beta_1 \\
  \alpha_2
 \end{ytableau} \,,
\end{equation}
which means that it satisfies
\begin{equation}
\phi_{[\alpha_1 \alpha_2] \beta_1} =  \phi_{\alpha_1\alpha_2  \beta_1}, \quad \phi_{[\alpha_1 \alpha_2 \beta_1]} = 0 \, .
\end{equation}
The gauge symmetries are given by  \cite{Curtright:1980yk}
\begin{align} \label{eq:phigauge}
\delta \phi_{\alpha_1 \alpha_2 \beta_1} = \pd_{[\alpha_1}S_{\alpha_2 ]\beta_1} +\pd_{[\alpha_1}A_{\alpha_2 ]\beta_1}- \pd_{\beta_1} A_{\alpha_1 \alpha_2}  \, ,
\end{align}
where $S$ and $A$ are arbitrary symmetric and antisymmetric tensors, respectively.
The corresponding gauge invariant curvature,
which we call the ``Riemann tensor'' $R$,
is defined by 
\begin{equation}
 R_{\alpha_1 \alpha_2 \alpha_3 \beta_1 \beta_2} 
\equiv 
\pd_{[\alpha_1}\phi_{\alpha_2 \alpha_3][ \beta_1, \beta_2]}
= \frac{1}{12} \pd_{\alpha_1}\!\pd_{\beta_2}\phi_{\alpha_2 \alpha_3 \beta_1} \pm \cdots
\end{equation}
It is a tensor of Young symmetry type $(2,2,1)$, i.e.\footnote{When there is no risk of confusion, we will omit the indices of the tableaux in the following.}, 
\begin{equation}
   R_{\alpha_1 \alpha_2 \alpha_3 \beta_1 \beta_2}
   \sim 
\begin{ytableau}
  \alpha_1 & \beta_1 \\
  \alpha_2 & \beta_2 \\
  \alpha_3
 \end{ytableau}
\end{equation}
which means that it satisfies
\begin{equation}\label{eq:symR}
R_{\alpha_1 \alpha_2 \alpha_3 \beta_1 \beta_2 }= R_{[\alpha_1 \alpha_2 \alpha_3] \beta_1 \beta_2 }= R_{\alpha_1 \alpha_2 \alpha_3 [\beta_1 \beta_2]}\, , \qquad R_{[\alpha_1 \alpha_2 \alpha_3 \beta_1] \beta_2 } = 0 \, .
\end{equation}
Its definition in terms of $\phi$ also implies that it satisfies the differential Bianchi identities
\begin{align}\label{Bianchi}
\pd_{[\mu}  R_{\nu \rho \sigma]\alpha \beta} = 0\, ,  \qquad R_{\mu \nu \rho [\alpha \beta , \gamma]} =0 \, .
\end{align}

The equations of motion for the chiral gauge field $\phi$,  $R = \hs R$, are then given by the self-duality condition on the
Riemann tensor on the first group of indices,
\begin{equation}
  R_{\alpha_1 \alpha_2 \alpha_3 \beta_1 \beta_2}
  =
  \frac{1}{3!}\varepsilon_{\alpha_1 \alpha_2 \alpha_3 \gamma_1 \gamma_2 \gamma_3}R\indices{^{\gamma_1 \gamma_2 \gamma_3 }_{\beta_1 \beta_2}} \,.
\label{self-duality}
\end{equation}
This condition makes sense due to the fact
that $\hs^2=1$ in that case. Because of the cyclic identity satisfied by $R$ (last of \eqref{eq:symR}), this condition implies the usual equation of motion
\begin{equation}
R\indices{^{\mu\nu\rho}_{\sigma\rho}} = 0 
\end{equation}
for a non-chiral $(2,1)$ field, but is stronger and only half the number of degrees of freedom are propagating ($8$ instead of $16$).

The question is to write an action principle yielding the equations \eqref{self-duality} or an equivalent subset.



\subsection{Electric and magnetic fields}
\label{sec:electr-magn-field}

In this subsection, we show how the equation $R = \hs R$ can be replaced by equations involving the spatial components $\phi_{ijk}$ only.

\paragraph{Definitions.}

We define the electric field of $\phi$ by
 \begin{equation}
 \mathcal{E}^{ijklm} \equiv \frac{1}{2! 3!} R_{pqrab} \varepsilon^{pqrlm}\varepsilon^{abijk}\, .
 \end{equation}
 It contains only the purely spatial components $\phi_{ijk}$. The electric field has Young symmetry {\tiny $\ydiagram{2,2,1}$}. It is also identically transverse in both groups of indices, i.e., $\pd_i{\mathcal E}^{ijklm}=0$ and $\pd_l{\mathcal E}^{ijklm}=0$.

The magnetic field is defined by the components of the curvature tensor with only one $0$,
 \begin{equation}
 {\mathcal B}_{ijklm} \equiv \frac{1}{2!} R\indices{_{0lm}^{ab}} \varepsilon_{abijk}\, .
 \end{equation}
It is identically double-traceless, $\bbar{{\mathcal B}}_{i}\equiv {\mathcal B}\indices{_{ilm}^{lm}}=0$, and transverse on the first group of indices, $\pd^{i}{\mathcal B}_{ijklm}=0$.

\paragraph{First step.}

The self-duality equation \eqref{self-duality} implies $\mathcal{E} = \mathcal{B}$,
\begin{equation}
{\mathcal E}^{ijklm} - {\mathcal B}^{ijklm}=0\, .
\label{E=B}
\end{equation}
Conversely, even though it does not contain the $\phi_{i00}$ variables, \eqref{E=B} implies all the components of the self-duality equation \eqref{self-duality} (see Appendix \ref{app:dem1}).
As a consequence of the double-tracelessness of the magnetic field, equation \eqref{E=B} implies that the electric
field is double-traceless on-shell, i.e.,
\begin{equation}
\bbar{\mathcal E}^{i} \equiv{\mathcal E}\indices{^{ilm}_{lm}}=0 \, .
\label{dtraceE}
\end{equation}
Similarly, the magnetic field has the $(2,2,1)$ symmetry on-shell.

\paragraph{Second step.}

The equation \eqref{E=B} still contains the $\phi_{0jk}$ components. To get rid of those, we take the curl of \eqref{E=B} on the second group of indices,
\begin{equation}
\curl_2(\mathcal{E} - \mathcal{B}) \equiv \varepsilon_{abcpq}\pd^{a}({\mathcal E}\indices{_{ijk}^{bc}}-{\mathcal B}\indices{_{ijk}^{bc}})=0 \, .
\label{curlE=B}
\end{equation}
This equation needs to be supplemented by \eqref{dtraceE}
which is a consequence of \eqref{E=B} 
that contains only $\phi_{ijk}$ components.

As we show in Appendix \ref{app:dem2}, there is no loss of information in going from \eqref{E=B} to the system \eqref{dtraceE}, \eqref{curlE=B}. This system is therefore equivalent to the original equations $R = \hs R$.

\subsection{Prepotential and action principle}
\label{sec:chiralaction}

In order to construct an action for the chiral tensor, we solve the constraint \eqref{dtraceE} by introducing a prepotential $Z_{abcij}$ for $\phi_{ijk}$, as was originally done for linearized gravity in~\cite{Henneaux:2004jw}.
Explicitly, the field can be written as
\begin{align}
\phi_{ijk} &= {\frac{1}{12}}{\mathbb P}_{(2,1)} \left( \partial^a Z\indices{^{bcd}_{ij}} \varepsilon_{kabcd} \right) + \text{(gauge)} \nonumber \\
&= \frac{1}{18} \left( \partial^a Z\indices{^{bcd}_{ij}} \varepsilon_{kabcd} - \partial^a Z\indices{^{bcd}_{k[i}} \varepsilon_{j]abcd} \right) + \text{(gauge)} \, .
\label{phiofZ}
\end{align}
The prepotential $Z_{abcij}$ has the $(2,2,1)$ Young symmetry, $Z \sim {\tiny \ydiagram{2,2,1}}\,$, i.e.,
\begin{equation}
Z_{abcij} = Z_{[abc]ij} = Z_{abc[ij]}\, , \quad Z_{[abci]j} = 0 \, .
\end{equation}
The fact that this expression solves the constraint is easily checked. Moreover, if $\phi$ is determined up to the gauge transformations \eqref{eq:phigauge}, the prepotential is determined up to the gauge and Weyl symmetries
\begin{equation}
\delta Z\indices{^{abc}_{ij}} = \delta_{g}Z\indices{^{abc}_{ij}} + \delta_{w}Z\indices{^{abc}_{ij}}\, ,
\label{invZ}
\end{equation}
where
\begin{align}
\delta_{g}Z\indices{^{abc}_{ij}} &= \pd^{[a}\alpha\indices{^{bc]}_{ij}}+ \pd^{[a}\beta\indices{^{bc]}_{[ij]}}- \frac{2}{3}\beta\indices{^{abc}_{[i,j]}} = {\mathbb P}_{(2,2,1)}\left( \frac{4}{3} \pd^{a}\alpha\indices{^{bc}_{ij}} + \pd_i \beta\indices{^{abc}_{j}}\right)\, , \label{eq:gaugeZ} \\
\delta_{w}Z\indices{^{abc}_{ij}} &= \rho^{[a}\delta^{bc]}_{ij}  = {\mathbb P}_{(2,2,1)}\left(\frac{4}{3} \rho^{a}\delta^{bc}_{ij}\right)\, , \label{eq:WeylZ}
\end{align}
as is again easily checked by direct substitution. Here, $\alpha\indices{^{ab}_{cd}}$, $\beta\indices{^{abc}_{d}}$ are $(2,2)$ and $(2,1,1)$ tensors, respectively, and the vector $\rho^{a}$ parametrizes Weyl rescalings.

The tensor that is invariant under the gauge transformations \eqref{eq:gaugeZ} is the Einstein tensor
\begin{equation}
    G\indices{_{de}^{l}}[Z] \equiv \frac{2}{3!4!}\varepsilon\indices{^{l}_{spqr}}\varepsilon_{deijk}\pd^{s}\pd^{i}Z^{pqrjk}\, .
\end{equation}
It is not invariant under the Weyl transformations \eqref{eq:WeylZ}. The invariant tensor controlling this Weyl invariance is
\begin{equation}\label{eq:defDtext}
    D_{abcde}[Z] \equiv \frac{1}{2}\varepsilon_{abclm}\pd^{m} S\indices{_{de}^{l}}[Z]\, ,
\end{equation}
which is called the Cotton tensor by analogy with the case of three dimensional gravity. In \eqref{eq:defDtext}, $S\indices{_{de}^{l}}[Z]$ is the Schouten tensor, defined from the Einstein by
\begin{equation}\label{eq:SofGtext}
S\indices{_{de}^{l}}[Z] \equiv G\indices{_{de}^{l}}[Z] + \frac{2}{3} \delta^{l}_{[d}G\indices{_{e]p}^p}[Z] \, .
\end{equation}
The construction and properties of these tensors are collected in Appendix \ref{sec:conf}.

Uniqueness of the formula \eqref{phiofZ} for the field $\phi$ in terms of the prepotential $Z$ then follows from the theorem of Appendix \ref{app:th2} about the Cotton tensor (conformal Poincaré lemma). Indeed, because the electric field $\mathcal{E}\indices{_{ijk}^{bc}}$ satisfies the three properties 
\begin{itemize}
    \item It is of Young symmetry type $(2,2,1)$,
    \item It is transverse in both groups of indices, 
    \item It is double-traceless (constraint),
\end{itemize}
there exists a field $Z\indices{^{pqrjk}}$ such that
\begin{equation}\label{eq:E=DZ}
\mathcal{E}\indices{_{abc}^{ij}} = D\indices{_{abc}^{ij}}[Z]\, .
\end{equation}
Equation \eqref{phiofZ} is such that \eqref{eq:E=DZ} is satisfied; moreover, it is uniquely determined from this condition up to gauge and Weyl transformations.

In addition, this implies for the curl of the magnetic field
\begin{equation}
    \frac{1}{2}\varepsilon_{abcpq}\pd^{a}{\mathcal B}\indices{_{ijk}^{bc}}[\phi[Z]]= \dot{D}_{ijkpq}[Z] \,.
\end{equation}
In terms of the prepotential $Z$, equation \eqref{curlE=B} therefore reads
\begin{equation}
\frac{1}{2}\varepsilon_{abcpq}\pd^{a}D\indices{_{ijk}^{bc}}\left[Z\right]-\dot{D}_{ijkpq}\left[Z\right]=0\, .
\label{eomZ}
\end{equation}
This equation follows from the variation of the action
\begin{equation}
    S[Z] = \frac{1}{2}\int \!dt\,d^5\!x\, Z_{ijkpq}\left(\dot{D}^{ijkpq}[Z] - \frac{1}{2} \varepsilon^{abcpq}\pd_{a}D\indices{^{ijk}_{bc}}\left[Z\right]  \right) \, .
    \label{actionZ}
\end{equation}

The action  (\ref{actionZ}) for the chiral field of $(2,1)$ Young symmetry type constitutes the central result of this section. It provides a variational principle for the original self-duality equations of motion $R = \hs R$, see \eqref{self-duality}.  The action is first order in the time derivatives of the prepotentials, which is the characteristic feature of an action in Hamiltonian form.  One can relate the chiral action derived here to the action for a non-chiral  field of $(2,1)$ Young symmetry type \cite{Curtright:1980yk}.  This is done in Appendix \ref{app:ham}.  By going to the Hamiltonian formulation and combining the Curtright field and its conjugate momentum -- or rather, the corresponding prepotentials which have  $(2,2,1)$ Young symmetry type -- into chiral and anti-chiral components, one gets the sum of the action (\ref{actionZ}) and the action for the anti-chiral component, which has the same structure but with the opposite sign for the kinetic term. Since these actions are decoupled, one can consistently drop the anti-chiral part, getting thereby the action (\ref{actionZ}) for a chiral field. 

Because the decomposition into chiral and anti-chiral components is covariant, the action (\ref{actionZ}) {\em is} covariant.  It is also intrinsically Hamiltonian, in the sense that there is no natural split of the prepotential $Z$ into $q$'s and $p$'s. One cannot therefore naturally eliminate half of the variables (``conjugate momenta'') to go to a second order formulation.  In order to be able to do so, one would need to keep the anti-chiral part.   It is the intrinsic Hamiltonian structure of the action (\ref{actionZ}) that is responsible for the fact that covariance, although present, is not manifest.  But this feature is just standard in the Hamiltonian formulation of relativistic field theories.

\section{The action of the free \texorpdfstring{$\CMcal{N} = (3,1)$}{N = (3,1)} theory}
\label{sec:Action}

We now have all the necessary tools to write down the action of the free $\cN = (3,1)$ theory.  The  fields  are displayed in Table \ref{Table1}, where their transformations properties both under the little algebra $\mathfrak{su}(2) \oplus \mathfrak{su}(2)$ and under the $R$-symmetry $\mathfrak{usp}(6) \oplus \mathfrak{usp}(2)$, as well as reality properties, are also given\footnote{We follow the conventions of \cite{Henneaux:2017xsb} for the $\mathfrak{usp}(2N)$ algebra and gamma matrices.}.  The bosonic spectrum contains one exotic chiral field of Young symmetry type $(2,1)$, $12$ chiral $2$-forms, $14$ vectors and $28$ scalars.  The fermionic spectrum contains $2$ exotic gravitini, $6$ standard chiral gravitini (Rarita-Schwinger fields), as well as $28$ left-handed and $14$ right-handed spin-$\frac12$-fields.

\begin{table}
\centering
$
\arraycolsep=5pt
\begin{array}{l l | l  c c c}
  \toprule
 \multicolumn{2}{c|}{\mathfrak{su}(2) \oplus \mathfrak{su}(2)}         & \multicolumn{1}{c}{\multirow{2}*{\text{Reality}}} & \multicolumn{1}{c}{\multirow{2}*{$\mathfrak{usp}(6)$\text{ irr.}}}                                                                                                                        & \multicolumn{1}{c}{\multirow{2}*{\text{Chirality}}}                                                                                                                                                   & \multicolumn{1}{c}{\multirow{2}*{\text{Dim.}}} \\ 
 \multicolumn{2}{c|}{\oplus\, \mathfrak{usp}(6)\oplus \mathfrak{usp}(2)} &                                                                                                                                                                                                                                              &                                                  &                                                                                                                                                                                                       & \\ \otoprule %
  (4,2;1,1)                                     & {\tiny\ydiagram{2,1}}                            & (Z\indices{^{ijk}_{pq}})^{*}=Z\indices{^{ijk}_{pq}}                                                                                                                            & -                                                                     &                   -                          & 1                                           \\ \midrule
  (3,1;6,2)                                     & {\tiny\ydiagram{1,1}}                           & A^{*}_{\tta \alpha ij} =    \Omega_{\tta \ttb} \varepsilon_{\alpha\beta}  A^{\ttb \beta}_{ij}                                                                                                         & -                                                                                &                                          -    & 2                                           \\ \midrule
  \multirow{2}*{$(2,2;14,1)$}                   & \multirow{2}*{{\tiny\ydiagram{1}}}                           & V_{\tta \ttb i}^{*} =  \Omega_{\tta \ttc} \Omega_{\ttb \ttd} V^{\ttc \ttd}_{i}                                                                                                                              & \Omega_{\tta \ttb} V^{\tta \ttb}=0                                                                        &                                           -   & \multirow{2}*{2} \\ 
                                                &                                                              & W_{\tta \ttb ijk}^{*} =  \Omega_{\tta \ttc} \Omega_{\ttb \ttd} W^{\ttc \ttd}_{ijk}                                                                                                                          & \Omega_{\tta \ttb} W^{\tta \ttb}_{ijk}=0                       & - \\ \midrule
 \multirow{2}*{$(1,1;14^{\prime} ,2)$}          & \multirow{2}*{$\bullet$}                                     & \phi^{*}_{\tta \ttb \ttc \alpha} =  \Omega_{\tta \tta^{\prime}} \Omega_{\ttb \ttb^{\prime}}\Omega_{\ttc \ttc^{\prime}} \varepsilon_{\alpha \alpha^{\prime}}\phi^{\tta^\prime \ttb^\prime \ttc^\prime \alpha^{\prime}} & \Omega_{\tta \ttb} \phi^{\tta \ttb \ttc \alpha}=0                   &          -        & 2 \\
                                                &                                                              & \pi^{*}_{\tta \ttb \ttc \alpha} =  \Omega_{\tta \tta^{\prime}} \Omega_{\ttb \ttb^{\prime}}\Omega_{\ttc \ttc^{\prime}} \varepsilon_{\alpha \alpha^{\prime}} \pi^{\tta^\prime \ttb^\prime \ttc^\prime \alpha^{\prime}}  & \Omega_{\tta \ttb} \pi^{\tta \ttb \ttc \alpha}=0                                                                          &                                          -    & 3 \\ \otoprule
  (3,2;6,1)                                     & {\tiny\ydiagram{1}}_{\text{F}}                               & \theta^{*}_{\tta ijk} = \Omega_{\tta \ttb} \mcB \theta^{\ttb}_{ijk}                                                                                                                                   &  -                                                                      & \Gamma_{7}\theta=+\theta                      & \frac{3}{2} \\ \midrule
  (4,1;1,2)                                     & {\tiny\ydiagram{1,1}}_{\text{F}}                          & \chi^{*}_{\alpha ij} =\varepsilon_{\alpha\beta} \mcB \chi_{ij}^{\beta}                                                                                                            &     -                                               & \Gamma_{7}\chi=+\chi                          & \frac{3}{2} \\ \midrule
  (2,1;14,2)                                    & \bullet_{\text{F}}                                           & \psi^{*}_{\tta \ttb \alpha} =  \Omega_{\tta \tta^{\prime}} \Omega_{\ttb \ttb^{\prime}} \varepsilon_{\alpha \alpha^{\prime}} \mcB \psi^{\tta^\prime \ttb^\prime  \alpha^{\prime}}                    &  \Omega_{\tta \ttb} \psi^{\tta \ttb \alpha}=0                                                                                                  & \Gamma_{7}\psi=+\psi                          & \frac{5}{2} \\ \midrule
  (1,2;14^{\prime},1)                           & \bullet_{\text{F}}                                           & \widetilde{\psi}^{*}_{\tta \ttb \ttc} =  \Omega_{\tta \tta^{\prime}} \Omega_{\ttb \ttb^{\prime}}\Omega_{\ttc \ttc^{\prime}}  \mcB \widetilde{\psi}^{\tta^\prime \ttb^\prime \ttc^\prime}                                & \Omega_{\tta \ttb} \widetilde{\psi}^{\tta \ttb \ttc}=0                                                  & \Gamma_{7}\widetilde{\psi}=-\widetilde{\psi} & \frac{5}{2} \\ \bottomrule
\end{array}
$
\caption{The fields of the $\cN = (3,1)$ theory and their transformation properties. We have indicated the space-time transformation properties by the corresponding Young diagram (with an extra $F$ index in the case of fermions). Indices $\tta, \ttb = 1,..,6$ and $\alpha, \beta = 1, 2$ label the fundamental representations of $\mathfrak{usp}(6)$ and $\mathfrak{usp}(2)$, respectively. Quantities with multiple indices transform
in the corresponding tensor product.  We also indicate the reality, irreducibility and (in the case of fermions) chirality conditions they satisfy. In the fermionic reality conditions, the $\mcB$ matrix is defined by the complex conjugation property $(\Gamma_\mu)^* = \mcB \Gamma_\mu \mcB^{-1}$ and is given by $\mcB = -i C \Gamma^0$ in terms of the usual charge conjugation matrix $C$. Canonical dimension is indicated in the last column.}
\label{Table1}
\end{table}

The action is a sum of eight terms, one for each type of fields,
\begin{equation}
    S= S_{{\tiny\ydiagram{2,1}}}+ S_{{\tiny\ydiagram{1,1}}}+S_{1} +S_{0}+ S^{L}_{{\tiny\ydiagram{1,1}}_{\text{F}}}+S^{L}_{3/2}+S^{L}_{1/2} + S^{R}_{1/2} \, .
\end{equation}

\subsection{Bosonic fields}

\begin{itemize}

\item{Chiral $(2,1)$-tensor }

This is the real tensor field of mixed Young symmetry $(2, 1)$ type with self-dual field strength described in the previous section. The action is written in terms of the prepotential $Z_{ijklm}$; it was derived in subsection \ref{sec:chiralaction} and reads
\begin{equation}
    S_{\tiny\ydiagram{2,1}}[Z] = \frac{1}{2}\int \!d^6 \!x\, Z_{ijklm}\left(\dot{D}^{ijklm}[Z] - \frac{1}{2} \varepsilon^{abclm}\pd_{a}D\indices{^{ijk}_{bc}}\left[Z\right]  \right)        \, , 
\end{equation}
where the Cotton tensor $D^{ijklm}[Z]$ is explicitly given in equation \eqref{eq:defDtext}.

\item{Chiral 2-forms}

The theory contains $12$ chiral 2-forms $A^{\tta\alpha}_{ij}$, described by the action \cite{Henneaux:1988gg}
\begin{equation}
S_{\tiny\ydiagram{1,1}}[A] = - \frac{1}{2}\int \!d^6 \!x\, A_{\tta\alpha ij}^{*} \left( \dot \cB^{\tta\alpha ij}[A]- \frac{1}{2} \varepsilon^{ijklm} \pd_{k} \cB^{\tta\alpha}_{lm}[A] \right)   \, ,
\end{equation}
where the magnetic fields $\cB^{\tta \alpha}_{ij}$ are given by
\begin{equation}
    \cB^{\tta \alpha i j}[A] = \frac{1}{2}\varepsilon^{ijklm}\pd_{k}A^{\tta \alpha}_{lm} \, .
\end{equation}

\item{Vector fields}

The theory possesses 14 vector fields $V_\mu^{\tta\ttb}$.  We write the action in Hamiltonian form.  The conjugate momenta $\Pi^m_{\tta\ttb}$ to the spatial components $V_m^{\tta\ttb}$ of the vector potentials are subject to Gauss' law $\partial_m \Pi^m_{\tta\ttb} \approx 0$, enforced by the temporal components $V_0^{\tta\ttb}$. This constraint can be solved by introducing a $3$-form potential $W^{\tta \ttb}_{ijk}$ such that the action takes the form \cite{Schwarz:1993vs,Bunster:2011qp}
\begin{align}
S_1[V,W] =- \frac{1}{2} \int \!d^6 \!x\, \Big[ &V^{*}_{\tta \ttb i} \left( \dot \cB^{\tta \ttb i}[W] + \frac{1}{3!} \varepsilon^{ijklm} \pd_{j} \cB^{\tta \ttb}_{klm}[V] \right) \nonumber \\
& \left.+\frac{1}{3!} W^{*}_{abijk} \left(-\dot{\cB}^{\tta \ttb ijk}[V] +\epsilon^{ijklm}\pd_{l} \cB^{\tta \ttb}_{m}[W] \right)\right]\, ,
\end{align}
where the magnetic fields are
\begin{align}
    \cB^{\tta \ttb ijk}[V] = \varepsilon^{ijklm}\pd_{l} V^{\tta \ttb}_{m} \, , \quad \cB^{\tta \ttb i}[W] = \frac{1}{3!}\varepsilon^{ijklm}\pd_{j} W^{\tta \ttb}_{klm} \, . 
\end{align}

\item{Scalar fields}

The action for the 28 scalar fields $\phi^{\tta \ttb \ttc \alpha}$ can be written in Hamiltonian form in the following way:
\begin{equation}
    S_0[\phi,\pi] = \frac{1}{2} \int \!d^6 \!x\, \left(2 \,\pi^{*}_{\tta \ttb \ttc \alpha}  \dot{\phi}^{\tta \ttb \ttc \alpha} -\pi_{\tta \ttb \ttc \alpha}^{*} \pi^{\tta \ttb \ttc \alpha} - \pd_{i} \phi_{\tta \ttb \ttc \alpha}^{*} \pd^{i} \phi^{\tta \ttb \ttc \alpha} \right)  \, .
\end{equation}

\end{itemize}

\subsection{Fermionic fields}

\begin{itemize}

\item{Exotic gravitini}

The $2$ left-handed exotic gravitini are described by fermionic $2$-form prepotentials $\chi^{\alpha}_{ij}$. The action in terms of prepotentials was derived in \cite{Henneaux:2017xsb} and reads
\begin{equation}
S^L_{{\tiny\ydiagram{1,1}}_{\text{F}}}[\chi] = -2 i \int \!d^6 \!x\, \chi^{\dagger}_{\alpha ij} \left(\dot D^{\alpha ij}[\chi] - \frac{1}{2}\varepsilon^{ijklm} \pd_{k} D^{\alpha}_{lm}[\chi] \right). 
\end{equation}
Here, the Cotton tensor is defined as
\begin{equation}
D^{\alpha ij}[\chi] = \varepsilon^{ijklm}\pd_{k} S^\alpha_{lm}[\chi] \, , \end{equation}
where the Schouten tensor is
\begin{equation}
    S^\alpha_{ij}[\chi] = - \left(\delta^{[k}_{[i} \Gamma\indices{_{j]}^{l]}} + \frac{1}{6} \Gamma_{ij} \Gamma^{kl} \right) \varepsilon_{klpqr}\pd^p \chi^{\alpha\, qr}\, .
\end{equation}

\item{Gravitini}

The theory contains $6$ left-handed gravitini, described by the prepotentials $\theta^\tta_{ijk}$. The action is \cite{Lekeu:2018kul}
\begin{equation}\label{eq:actiongravitini}
    S^L_{3/2}[\theta] = -i \int \!d^6 \!x\, \theta^{\dagger}_{\tta ijk} \left(\dot D^{\tta ijk}[\theta] - \varepsilon^{ijklm} \pd_{j} \widetilde{D}^{\tta}_{m}[\theta] \right) \, .
\end{equation}
The Cotton tensor is defined by
\begin{equation}
D^{\tta ijk}[\theta] = \varepsilon^{ijklm} \pd_{l}S^{\tta}_{m}[\theta] \, ,
\end{equation}
where the Schouten tensor $S^{\tta}_{i}$ is defined as
\begin{equation}
    S^{\tta}_{i}[\theta]=\frac{1}{4}\left( \Gamma_{ij} - 3 \delta_{ij}\right) \varepsilon^{jklmn}\pd_{k}\theta^{\tta}_{lmn} \,.
\end{equation}
In \eqref{eq:actiongravitini}, the quantity $\widetilde{D}^\tta_m$ is the gamma-matrix contraction
   \begin{equation}
           \widetilde{D}^\tta_{m} = \frac{1}{2} \Gamma^{ab} D^\tta_{mab} \, .
   \end{equation}

\item{Spin $1/2$ fields}

Finally, the theory possesses $28$ and $14$ left-handed and right-handed spin $1/2$ fields, which we write as $\psi^{\tta \ttb \alpha}$ and $\widetilde{\psi}^{\tta \ttb \ttc}$, respectively. Their action is simply
\begin{equation}
    S^{L}_{1/2}[\psi] = i \int \!d^6 \!x\, \psi^{\dagger}_{\tta \ttb \alpha} \left(\dot \psi^{\tta \ttb \alpha} - \Gamma^{0} \Gamma^{i} \pd_{i} \psi^{\tta \ttb \alpha} \right)
\end{equation}
and
\begin{equation}
    S^{R}_{1/2}[\widetilde{\psi}] = i \int \!d^6 \!x\, \widetilde{\psi}^{\dagger}_{\tta \ttb \ttc} \left(\dot{\widetilde{\psi}}{}^{\tta \ttb \ttc} - \Gamma^{0} \Gamma^{i} \pd_{i} \widetilde{\psi}^{\tta \ttb \ttc} \right) \, .
\end{equation}

\end{itemize}

\subsection{Comments on the \texorpdfstring{$F_{4(4)}$}{F 4(4)} symmetry}

As was noticed by Hull, ${USp(6) \times USp(2)}$ is the maximal compact subgroup of the exceptional group $F_{4(4)}$, and the number of scalar fields is the same as the dimension of the coset space
\begin{equation}\label{eq:F4coset}
\frac{F_{4(4)}}{USp(6) \times USp(2)} \, .
\end{equation}
By analogy with the well-known situation in extended supergravities, it is natural to conjecture that, in the interacting theory, the scalar fields indeed take values in \eqref{eq:F4coset} \cite{Hull:2000zn}.

The $F_{4(4)}$ symmetry should then also extend as a symmetry of the full bosonic sector. It is interesting to notice that the lowest-dimensional irreducible representation of $F_{4(4)}$ is exactly of dimension $26$: therefore, we conjecture that the $12$ chiral $2$-forms combine with the $14$ vector fields into an irreducible multiplet, while the chiral tensor is a singlet under $F_{4(4)}$. This is supported by the branching rules of representations for the embeddings
\begin{equation}
USp(6) \times USp(2) \,\subset\, F_{4(4)} \,\subset\, E_{6(6)}
\end{equation}
(see for example \cite{Yamatsu:2015npn} for tables):
\begin{enumerate}
\item Under its maximal compact subgroup $USp(6) \times USp(2)$, the fundamental representation of $F_{4(4)}$ decomposes as
\begin{equation}
\mathbf{26} = (6,2) \oplus (14,1),
\end{equation}
which are exactly the $USp(6) \times USp(2)$ transformation rules of the chiral forms and vector fields.
\item Under its $F_{4(4)}$ subgroup, the fundamental representation of $E_{6(6)}$ decomposes as
\begin{equation}
\mathbf{27} = \mathbf{26} \oplus \mathbf{1} .
\end{equation}
It is well known that the $27$ vector fields of maximal supergravity in five dimensions fall into an irreducible $E_{6(6)}$ multiplet \cite{Cremmer:1979uq}. From the dimensional reduction of the
$(3,1)$-theory, $26 = 12 + 14$ of those vector fields arise from the six-dimensional chiral forms and vector fields, and one comes from the reduction of the chiral $(2,1)$ tensor. Therefore, the well-known  $E_{6(6)}$ symmetry of five-dimensional maximal supergravity seems to support this conjecture.
\end{enumerate}
Of course, it is impossible to prove this in the absence of a consistent interacting theory. Moreover, this symmetry would be very peculiar since it would mix fields with different spacetime indices (see also \cite{Ananth:2017nbi} for a similar situation). Nevertheless, we feel that these remarks are intriguing and could have useful implications for the interacting theory, if it exists.

\section{Supersymmetry}
\label{sec:susy}

We now establish that the action of the previous section is invariant under $\cN = (3,1)$ supersymmetry. The supersymmetry parameters are $\epsilon^\tta$ and $\et^\alpha$. As their indices suggest, they transform in the fundamental of $\mathfrak{usp}(6)$ and $\mathfrak{usp}(2)$, respectively, and are inert under the other factor. They satisfy symplectic Majorana-Weyl conditions gathered in table \ref{tab:susy}, and their canonical dimension is $-1/2$.
\begin{table}[h]
    \centering
    $
\begin{array}{l | c c c}
  \toprule
             & \multicolumn{1}{c}{\text{Reality}}                         & \multicolumn{1}{c}{\text{Chirality}}     & \multicolumn{1}{c}{\text{Dimension}} \\ \midrule
  \epsilon^{\tta} \; & \;  \epsilon_{\tta}^{*}= \Omega_{\tta\ttb} \,\mcB\, \varepsilon^{\ttb}         \; & \; \Gamma_{7} \epsilon^{\tta} = -\epsilon^{\tta} & -1/2                         \\
 \et^{\alpha} \; & \; \et_{\alpha}^{*} = \varepsilon_{\alpha\beta} \,\mcB\, \et^{\beta} \; & \; \Gamma_{7}\et^{\alpha} = + \et^{\alpha} & -1/2                         \\ \bottomrule
\end{array}
$
    \caption{Supersymmetry parameters of the $(3,1)$-theory.}
    \label{tab:susy}
\end{table}

The supersymmetry transformations on the various fields are then
\begin{align}
\delta Z\indices{^{ijk}_{lm}} &=  \mathbb{P}_{(2,2,1)}\left( \beta_1\bar{\epsilon}_\tta \,\Gamma_{lm} \theta^{\tta ijk}  + \beta_{2} \bar{\epsilon}_{\alpha}\left(\Gamma^{ijk} \chi_{lm}^{\alpha}+12 \Gamma^{[i}\delta^{j}_{[l}\chi\indices{^{k]\alpha}_{m]}} \right)\right) \\
\delta\theta^{\tta}_{ijk} &= -\frac{\beta_1}{8 \cdot 3!^2} \left(\pd^r Z\indices{_{ijk}^{ab}}\varepsilon_{pqrab}\Gamma^{pq} \Gamma^{0}\epsilon^{\tta} + \frac{2}{3} \pd^{a} Z\indices{^{bcd}_{[ij}}\varepsilon_{k]abcd}\Gamma^{0}\epsilon^{\tta}\right)\nonumber \\
&\quad + \beta_{3}\left(\varepsilon_{\alpha \beta} A^{\tta \alpha}_{[ij}\Gamma_{k]}\Gamma^{0}\tilde{\epsilon}^{\beta}\right)+ \beta_{4}\left( V\indices{^{\tta\ttb}_{[i}}\Omega_{\ttb \ttc} \Gamma_{jk]} \Gamma^0 \epsilon^\ttc +\frac{1}{3} W^{\tta\ttb}_{ijk} \Omega_{\ttb\ttc} \Gamma^{0} \epsilon^\ttc\right) 
\end{align}
\begin{align}
\delta\chi^{\alpha }_{ij} &= \frac{\beta_{2}}{4\cdot 4!}\left(\varepsilon^{abcde}\pd_{e}Z_{abc ij} \Gamma_{d}\Gamma^{0}\tilde{\epsilon}^{\alpha} \right)+ \beta_{5}\left( \frac{1}{2} A^{\tta \alpha}_{ij}\Omega_{\tta\ttb}\Gamma^{0}\epsilon^{\ttb}\right) \\
\delta A^{\tta\alpha}_{ij} &=  \beta_{3}\left(-4 \varepsilon^{\alpha \beta}\bar{\tilde{\epsilon}}_{\beta } \Gamma_{[i} S^{\tta}_{j]}[\theta]\right)+\beta_{5} \left(4 \Omega^{\tta\ttb} \bar{\epsilon}_\ttb S^{\alpha}_{ij}[\chi] \right)+\beta_{6} \left(-2 \bar{\epsilon}_{\ttb} \Gamma_{ij} \psi^{\tta\ttb \alpha} \right) \\
\delta V^{\tta\ttb}_{i} &= - \beta_{4} \left(4  \Omega^{\ttc[\tta}\bar{\epsilon}_\ttc S^{\ttb]}_{i}[\theta]+\frac{2}{3} \Omega^{\tta\ttb} \bar{\epsilon_\ttc} S^\ttc_{i}[\theta]\right) \nonumber \\
&\quad+ \beta_{7} \left( 2 \bar{\tilde{\epsilon}}_\alpha \Gamma_{i} \psi^{\tta\ttb\alpha}\right) + \beta_{8}\left( 2 \bar{\epsilon}_\ttc  \Gamma_{i} \tilde{\psi}^{\tta\ttb\ttc}\right) \\ 
\delta W^{\tta\ttb}_{ijk} &= - \beta_{4} \left(12  \Omega^{\ttc[\tta} \bar{\epsilon}_\ttc \Gamma_{[ij} S^{\ttb]}_{k]}[\theta]+2\Omega^{\tta\ttb} \bar{\epsilon}_\ttc \Gamma_{[ij} S^{\ttc}_{k]}[\theta] \right) \nonumber \\
&\quad+ \beta_{7} \left( 2 \bar{\tilde{\epsilon}}_\alpha \Gamma_{ijk} \psi^{\tta\ttb\alpha}\right) + \beta_{8}\left(2 \bar{\epsilon}_\ttc \Gamma_{ijk} \tilde{\psi}^{\tta\ttb\ttc}\right) \\
\delta \psi^{\tta\ttb\alpha}&= \beta_{7} \left( \cB^{\tta\ttb}_{i}[W] \Gamma^i +\frac{1}{3!} \cB^{\tta\ttb}_{ijk}[V] \Gamma^{ijk} \right) \Gamma^{0} \tilde{\epsilon}^{\alpha} \nonumber \\
&\quad+\beta_{9}\left( \pi^{\tta\ttb\ttc\alpha} \Omega_{\ttc\ttd}\Gamma^0 \epsilon^{\ttd} +  \pd_i \phi^{\tta\ttb\ttc\alpha} \Omega_{\ttc\ttd} \Gamma^{i}\epsilon^{\ttd} \right) \nonumber \\
&\quad+\beta_{6} \Gamma^{ij} \Gamma^{0} \left(\cB^{[\tta|\alpha}_{ij}[A]  \epsilon^{|\ttb]}-\frac{1}{6}\Omega^{\tta \ttb} \cB^{\ttc \alpha}_{ij}[A] \Omega_{\ttc\ttd} \epsilon^{\ttd}\right) \\
\delta \tilde{\psi}^{\tta\ttb\ttc} &= \beta_{10}\left( \pi^{\tta\ttb\ttc\alpha} \varepsilon_{\alpha \beta} \Gamma^{0} \tilde{\epsilon}^{\beta} +  \pd_i \phi^{\tta\ttb\ttc\alpha} \Gamma^{i}\varepsilon_{\alpha \beta}\tilde{\epsilon}^{\beta}\right)\nonumber \\
&\quad+\beta_{8} \Gamma^i \Gamma^{0} \left( \cB^{[\tta\ttb}_{i}[W] \epsilon^{\ttc]} -\frac{1}{2}\Omega^{[\tta \ttb} \cB^{\ttc]\ttd}_{i}[W] \Omega_{\ttd \tte}\epsilon^{\tte} \right) \nonumber \\
&\quad +\frac{\beta_8}{3!} \Gamma^{ijk}\Gamma^{0}\left( \cB^{[\tta\ttb}_{ijk}[V] \epsilon^{\ttc]}-\frac{1}{2} \Omega^{[\tta \ttb}\cB^{\ttc]\ttd}_{ijk}[V] \Omega_{\ttd \tte} \epsilon^{\tte}\right) \\
\delta \pi^{\tta\ttb\ttc\alpha} &= \beta_{10} \left( -2\varepsilon^{\alpha \beta } \bar{\tilde{\epsilon}}_{\beta} \Gamma^{i} \Gamma^{0} \pd_i \tilde{\psi}^{\tta\ttb\ttc}\right)+ \beta_{9} \left(- 2 \bar{\epsilon}_{\ttd} \Gamma^{i} \Gamma^{0} \pd_i \psi^{[\tta\ttb| \alpha} \Omega^{\ttc]\ttd} - \bar{\epsilon}_{\ttd} \Gamma^{i} \Gamma^{0}  \Omega^{[\tta\ttb}\pd_i \psi^{\ttc]\ttd\alpha} \right) \\
\delta \phi^{\tta\ttb\ttc\alpha} &= \beta_{10} \left( 2 \varepsilon^{\alpha \beta} \bar{\tilde{{\epsilon}}}_\beta \tilde{\psi}^{\tta\ttb\ttc}\right) + \beta_{9} \left(2 \bar{\epsilon}_{\ttd} \psi^{[\tta\ttb |\alpha}\Omega^{\ttc]\ttd} + \bar{\epsilon}_\ttd \Omega^{[\tta\ttb} \psi^{\ttc]\ttd\alpha}  \right) \, . 
\end{align}
They were found from the following requirements:
\begin{enumerate}[a)]
\item They leave the action invariant;
\item Symplectic indices, reality and chirality conditions must match;
\item A gauge and Weyl transformation of the right hand-side must induce a gauge and Weyl transformation of the field of the left-hand side.
\end{enumerate}
Conditions b) and c) are actually sufficient to fix nearly all the variations. The remaining transformations (and relative factors) are then found from condition a). A consequence of condition c) is that the supersymmetry transformations of the invariant curvatures (Cotton tensors and magnetic fields) can be expressed purely with invariant objects. This is indeed what we find:
\begin{align}
\delta D^{\tta ijk}[\theta] & = -\frac{\beta_1}{2}D^{ijklm}[Z]  \Gamma_{lm} \Gamma^{0}\epsilon^{\tta} \nonumber \\
&\quad- \beta_3\left(\frac{1}{2}\varepsilon^{ijklm}\Gamma\indices{_{m}^{pq}} \pd_{l}\cB^{\tta \alpha}_{pq}[A]-2\epsilon^{ijklm}\pd_{l}\cB^{\tta \alpha}_{mp}[A] \Gamma^{p} \right) \Gamma^{0}\varepsilon_{\alpha \beta} \tilde{\epsilon}^{\beta} \nonumber \\
&\quad +\frac{\beta_4}{4} \left( \varepsilon^{ijklm} \pd_{l}\Gamma\indices{_{m}^{pqr}} \cB^{\tta \ttb}_{pqr}[V]- \varepsilon^{ijklm}\pd_{l}\Gamma^{pq}\cB^{\tta \ttb}_{mpq}[V] \right)\Gamma^{0}\Omega_{\ttb \ttc}\epsilon^{\ttc} \nonumber \\
&\quad +\frac{\beta_4}{2}\left(\varepsilon^{ijklm}\pd_{l} \Gamma\indices{_{m}^{p}} \cB^{\tta \ttb}_{p}[W] -3 \epsilon^{ijklm}\pd_{l}\cB^{\tta \ttb}_{m}[W] \right)\Gamma^{0}\Omega_{\ttb \ttc} \epsilon^{\ttc} \\
\delta D^{\alpha }_{ij}[\chi]&=\frac{\beta_1}{4}\Gamma^{0}\left(  D\indices{^{klm}_{ij}}[Z]\Gamma_{klm} +12 D\indices{^{k}_{[i}^{l}_{j]l}}[Z]\Gamma_{k}\right)\tilde{\epsilon}^{\alpha} \nonumber \\
&\quad + \frac{\beta_5}{3}\left(\varepsilon_{ijklm} \pd^{k} \cB^{\tta \alpha lm}[A] + \varepsilon_{ijklm} \pd^{k} \Gamma\indices{^{l}_{p}} \cB^{\tta \alpha mp}[A] + 2 \Gamma_{k}\Gamma_{0} \pd^{k} \cB^{\tta \alpha}_{ij}[A]\right)\Omega_{\tta \ttb}\Gamma^{0}\epsilon^{\ttb}
\end{align}
\begin{align}
\delta D\indices{^{ijk}_{lm}}[Z] &= \frac{\beta_1 }{4 \cdot 3!^2} \bar{\epsilon}_\tta \mathbb{P}_{(2,2,1)} \left( \varepsilon_{pqrlm}\Gamma^{pq} \pd^r D^{\tta ijk}[\theta] + \frac{2}{3} \varepsilon^{ijkpq} \pd_p D^{\tta}_{ qlm}[\theta] \right) \\
&\quad +  \frac{\beta_2}{24} \bar{\epsilon}_{\alpha}\mathbb{P}_{(2,2,1)}\left( \varepsilon^{ijkpq}\Gamma_{p}\pd_{q} D^{\alpha}_{lm}[\chi]\right) \\
\delta \cB^{\tta\alpha}_{ij}[A] &= \beta_{6} \left(-\bar{\epsilon}_{\ttb} \varepsilon_{ijklm} \pd^k \Gamma^{lm} \psi^{\tta\ttb \alpha} \right)+ \beta_{5} \left(-2\Omega^{\tta\ttb}\bar{\epsilon}_{\ttb}D^{\alpha}_{ij}[\chi] \right)+\beta_{3}\left(2 \varepsilon^{\alpha \beta} \bar{\tilde{\epsilon}}_{\beta } \Gamma^{k} D\indices{^{\tta}_{ijk}}[\theta] \right)\, . 
\end{align}
The action is invariant for any values of the (real) constants $\beta_1$ to $\beta_{10}$. These constants are fixed by the supersymmetry algebra: they must satisfy
\begin{equation}\label{eq:relbetas}
\frac{\beta_1^2}{36} = \frac{\beta_2^2}{8} = 4 \beta_3^2 = 4 \beta_4^2 = 2 \beta_5^2 = 4 \beta_6^2 = 4 \beta_7^2 = \frac{4 \beta_8^2}{3} = \frac{2 \beta_9^2}{3} = 2 \beta_{10}^2 \equiv \kappa^2 .
\end{equation}
The proof is a bit technical and presented in Appendix \ref{app:susy}. The method we follow is dimensional reduction and comparison with linearized maximal supergravity in five dimensions \cite{Cremmer:1979uq}: this allows us to bypass the somewhat cumbersome direct computation of the $\cN = (3,1)$ commutators.

The supersymmetry algebra dictates  the equality of the numbers of bosons and fermions through the anticommutation relation $\{Q,Q\} \sim P$.  This equality holds only for the full spectrum, and not for any of the 10 individual actions obtained by dropping from the theory the fields that do not transform under the individual fermionic symmetries with one single non-vanishing $\beta_i$ ($i = 1, \cdots, 10$). Consequently, the fact that the constants $\beta_i$ are related by the supersymmetry algebra does not come as a surprise.

\section{Conclusions}
\label{sec:Conclu}

In this paper, we have constructed the action for the free $\cN = (3,1)$ theory of ``exotic supergravity'' in six spacetime dimensions.   This step fills the last box in Figure \ref{fig:oxidation}.  It is satisfying to see   that all higher dimensional parents of $D = 4$, $\cN = 8$ maximal supergravity exist at the free level.  

The action not only encodes the equations of motion, but it also yields the Poisson brackets between the dynamical fields.  The pre-symplectic form derived from the kinetic term of the action is degenerate because of the gauge symmetries.  The brackets of invariants can however straightforwardly be computed.  The procedure is direct and  shows for instance that the Poisson bracket of two Cotton tensor components $D^{ijkrs}$ (or, what is the same, two electric ($\equiv$ magnetic) field components)  is proportional to the third derivative of the Dirac delta function \cite{Bunster:2013oaa},
\begin{equation}
[D^{ijkrs}({\vec x}), D^{i'j'k'r's'}({\vec y})] \sim \partial^3 \delta({\vec x}, {\vec y}) \, .
\end{equation} 
The detailed computation will not be reported here since the only point we want to make at this stage  is that the prepotential $Z$ is at the same time a ``$q$'' and a ``$p$'' (self-conjugate), as it is standard in the description of chiral bosonic  fields \cite{Floreanini:1987as}, \cite{Henneaux:1988gg}.

Because the variables in the variational principle are self-conjugate, there is no natural way to eliminate half of them to go to a second-order action. The action is intrinsically ``Hamiltonian'', i.e., first-order in the time derivatives.  The situation, thus, is not that there is no action, but that there is no natural second-order action.  This is in line with the fact that the (independent) self-duality conditions are after all of the first order in the time derivatives.  The Hamiltonian structure of the action, following from self-duality,  makes covariance less transparent.  This phenomenon is not peculiar to the exotic supergravity  theory considered here, but generically holds for the Hamiltonian formulation of relativistic field theories.   We note that spacetime covariance can be controlled by Hamiltonian tools \cite{Dirac:1962aa,Schwinger:1963zza}.

To conclude, we are of course fully aware that the construction of consistent (most likely non local) interactions for the two exotic theories of supergravity  in six spacetime dimensions remains a challenge that must be overcome in order to have physical theories.  We feel, however, that  the question is still open and that these intriguing and potentially rich theories  deserve further study.

\acknowledgments

Various equations were checked using the xAct suite
(available at \url{http://xact.es})
and in particular the xTrans package \cite{Nutma:2013zea}.

We thank Amaury Leonard for discussions.  V.L. is a Research Fellow at the Belgian F.R.S.-FNRS. This work was 
partially supported by the ERC Advanced Grant ``High-Spin-Grav''
and by FNRS-Belgium (Convention FRFC
 PDR T.1025.14 and Convention IISN 4.4503.15).


\appendix

\section{Equations of motion for the chiral \texorpdfstring{$(2,1)$}{(2,1)}-tensor}
\label{app:A}

In this appendix, we provide the explicit proofs of the equivalence between the self-duality condition $R = \hs R$ and the alternative forms considered in section \ref{sec:electr-magn-field}. This relies on the generalized Poincaré lemmas of \cite{Olver_hyper,DuboisViolette:1999rd,DuboisViolette:2001jk,Bekaert:2002dt} for tensors of mixed Young symmetry.

\subsection{First step: \texorpdfstring{$R = \hs R \Leftrightarrow {\mathcal E}={\mathcal B}$}{R = star R equals E=B}}
\label{app:dem1}

The self-dual equations of motion $R = \hs R$, see \eqref{self-duality}, can be split according to
\begin{eqnarray}
R_{ijklm} &=& \frac{3}{3!} \varepsilon_{ijk0ab}R\indices{^{0ab}_{lm}} = -\frac{1}{2}\varepsilon_{ijkab}R\indices{^{0ab}_{lm}}\, ,  \label{eqa}\\
R_{0 ij  kl} &=& \frac{1}{3!}\varepsilon_{ij abc}R\indices{^{abc }_{kl}} \, , \label{eqb} \\
R_{0ij0k}&=&\frac{1}{3!}\varepsilon_{ijabc}R\indices{^{abc}_{0k}} \, . \label{eqc}
\end{eqnarray}
Due to the Young symmetries of the Riemann tensor
the component $R_{abc0d}$ is not independent.
Explicitly, we can use the identity $R_{[\mu \nu \rho\sigma]\tau}=0$ to show $R_{\mu \nu \rho \sigma \tau}= 3 R_{\sigma[\mu \nu  \rho] \tau}$, from which
\begin{equation}
  \label{eq:Rconn}
   R_{abc 0d}=3 R_{0[abc]d}
\end{equation}
follows.
Contracting \eqref{eqb} with $\frac{1}{2!}\varepsilon^{klpqr}$
and
using the definitions of the electric and magnetic fields
we obtain
\begin{align}\label{eq:BEfoll}
  \frac{1}{2!}\varepsilon^{klpqr}R_{0 ij kl}
  =
      \frac{1}{2!3!}\varepsilon^{klpqr}\varepsilon_{ij abc}R\indices{^{abc}_{kl}}
  \qquad \Rightarrow \qquad 
  {\mathcal B}^{pqrij} ={\mathcal E}^{pqrij} \,.
\end{align}
So, ${\mathcal E}={\mathcal B}$ follows from $R = \hs R$.

In order to prove the converse, we must show that ${\mathcal E}={\mathcal B}$
implies \eqref{eqa}, \eqref{eqb} and \eqref{eqc}. 
Equation \eqref{eqb} follows from reversing the just given argument.
It is also easy to check that \eqref{eqa} is actually
equivalent to \eqref{eqb}. We contract \eqref{eqa} by
$\frac{1}{3!}\varepsilon^{ijkpq}$ such that
\begin{equation}
\frac{1}{3!}\varepsilon^{ijkpq}R_{ijklm} = -\frac{1}{2}\frac{2!3!}{3!}\delta^{pq}_{ab} R\indices{^{0ab}_{lm}}
= R\indices{_{0}^{pq}_{lm}} \,.
\end{equation}
Finally it remains to show that \eqref{eqc} follows from \eqref{eq:BEfoll} or, equivalently, \eqref{eqb}.
The problem is that \eqref{eqb} does not contain the $\phi_{i00}$ components, while \eqref{eqc} does. Those components appear in \eqref{eqc} with two spatial derivatives, i.e., in the form $\pd_{k}\pd_{[i}\phi_{j]00}$ (up to some factor that will not matter).

We will need to use the generalized Poincaré lemma of \cite{Bekaert:2002dt}
for a tensor $T_{ijk} \sim \ydiagram{2,1}$.
It states that, if the curl 
on the first and second set of indices vanishes,
then the tensor $T_{ijk}$  has to be proportional to
$\pd_{k}\pd_{[i}\lambda_{j]}$ for some vector $\lambda_j$.
Symbolically, this can be written as
\begin{equation}\label{eq:poincarelemma}
\ytableaushort{{}{},{}\pd}
= 0 , \quad
\ytableaushort{{}{},{},\pd} = 0 \quad \Rightarrow \quad \ydiagram{2,1} \sim \ytableaushort{{}\pd,\pd} \, .  
\end{equation}
The role of $T_{ijk}$ will be played by \eqref{eqc}
\begin{align}
  T_{ijk} =R_{0ij0k}-\frac{1}{3!}\varepsilon_{ijabc}R\indices{^{abc}_{0k}}\, ,
\end{align}
which indeed has the right Young symmetries. It is now sufficient to show that \eqref{eqb} implies that $T_{ijk}$ satisfies the assumptions of \eqref{eq:poincarelemma}: we then recover \eqref{eqc} up to a term of the form $\pd_{k}\pd_{[i}\lambda_{j]}$ which can be absorbed in a redefinition of $\phi_{i00}$.

Let us first show that the curl of $T_{ijk}$ on the second group of indices vanishes.
For that 
we take the time derivative of \eqref{eqb} which leads to
\begin{eqnarray}
\pd_0 R_{0ijkl}&=& \frac{1}{3!}\varepsilon_{ijabc} \pd_0 R\indices{^{abc}_{kl}} \,.
\end{eqnarray}
Next we use a consequence of the Bianchi identity~\eqref{Bianchi},
$R_{0ij kl ,0} = -2 R_{0ij0[k,l]}$,
to show that 
\begin{eqnarray}
R_{0ij0[k,l]}&=& \frac{1}{3!}\varepsilon_{ijabc} R\indices{^{abc}_{0[k,l]}}\, .
\end{eqnarray}
This is exactly the curl of \eqref{eqc} on the second group of indices.

Now, we would like to have a curl on the first group of indices.
For that, let us take again
the time derivative of \eqref{eqb}
and antisymmetrize the indices $[ijk]$
\begin{align}
 \pd_{0} R_{0 [ij k]l} =  \frac{1}{3!}\varepsilon_{abc[ij|}\pd_{0}R\indices{^{abc }_{|k]l}} \,.
\end{align}
On the left-hand side we
use \eqref{eq:Rconn} and 
a consequence of Bianchi identity, $\pd_0 R_{ijk 0l} = 3 \pd_{[i}R_{jk]00l}$.
The right hand side can also be worked out as
\begin{eqnarray}
 \varepsilon_{[ij | abc}\pd_{0}R\indices{^{abc }_{|k]l}} &=& \pd_{[i}\varepsilon_{jk]abc} R\indices{^{abc}_{0l}}- \pd_{l} \varepsilon_{[ij|abc} R\indices{^{abc}_{0|k]}} \, ,
\end{eqnarray}
where we have used again a Bianchi identity
$R_{ij klm ,0} = -2 R_{ijk0[l,m]}$.
It is easy to check that the second term on the right-hand side vanishes on-shell.
For that, we take \eqref{eqb}, antisymmetrize the indices $[ijk]$, and use \eqref{eq:Rconn} to get $R\indices{^{abc}_{0k}} = \frac{1}{2}\varepsilon^{[ab|def}R\indices{_{def}^{|c]}_{k}}$. 
We now plug this into the second term to show
\begin{equation}
  \varepsilon_{[ij|abc}R\indices{^{abc}_{0|k]}}
  \sim
  \varepsilon_{[ij|abc}\varepsilon^{abdef} R\indices{_{def}^{c}_{|k]}}
  \sim
  R\indices{_{[ij|c}^{c}_{|k]}}=0 \,. 
\end{equation}
The last equality holds because the trace $R\indices{_{ijc}^{c}_{k}}$ has the $(2,1)$ Young symmetry.
So we finally arrive at
\begin{eqnarray}
  \pd_{[i} R_{ j k]00l}
  &=&
      \frac{1}{3!}\pd_{[i}\varepsilon_{jk]abc} R\indices{^{abc}_{0l}}\,,
\end{eqnarray}
which is the curl of \eqref{eqc} on the first group of indices.

As anticipated, we can now use
the generalized Poincaré lemma of \cite{Bekaert:2002dt}
to get
\begin{equation}
\label{eq:eqcprop}
R_{0ij0k}-\frac{1}{3!}\varepsilon_{ijabc}R\indices{^{abc}_{0k}} =\pd_{k}\pd_{[i}\lambda_{j]} \,.
\end{equation}
The $\pd_{k}\pd_{[i}\lambda_{j]}$ terms can be absorbed
by redefining the $\phi_{i00}$ terms of $R_{0ij0k}$.

\subsection{Second step: \texorpdfstring{${\mathcal E}={\mathcal B}\Leftrightarrow \curl_{2}({\mathcal E}-{\mathcal B})=0$}{E = B equals curl(E-B)=0} and \texorpdfstring{$\bar{\bar{{\mathcal E}}}=0$}{doubletrace E = 0}}
\label{app:dem2}

Taking the curl
on the second pair of indices
of \eqref{E=B}
suffices to show that 
\begin{equation}\label{eq:curlEB}
\curl_{2}({\mathcal E}-{\mathcal B})=\varepsilon_{abcpq}\pd^{a}({\mathcal E}\indices{_{ijk}^{bc}}-{\mathcal B}\indices{_{ijk}^{bc}})=0 \,.
\end{equation}
The double-tracelessness
of the electric field, $\bar{\bar{{\mathcal E}}}=0$, follows
because the magnetic field ${\mathcal B}$ is identically double-traceless.

To prove the converse, we introduce the tensor
\begin{equation}\label{eq:defK}
K_{ab cd} = \varepsilon\indices{^{ijk}_{cd}}({\mathcal E}-{\mathcal B})_{ijkab}\,.
\end{equation}
We have to show from \eqref{eq:curlEB} that the tensor $K$ can be written as $K_{abcd} = \pd_{[a} M_{b][c,d]}$. Indeed, this is the way in which the missing components $\phi_{0ij}$ appear in \eqref{eq:defK}: this will therefore imply $\mathcal E = \mathcal B$ up to a redefinition of the $\phi_{0ij}$. Again, the proof of this fact relies on generalized Poincaré lemmas.

The symmetries of tensor $K$ are such that 
$K_{abcd}=K_{[ab][cd]}$ and, 
because  ${\mathcal E}$ and ${\mathcal B}$ are double-traceless,
$K_{[abcd]}=0$. 
Moreover, it follows from its definition and \eqref{eq:curlEB} that its curl on the first and second group of indices vanishes,
\begin{eqnarray}
\pd_{[i}K_{ab]cd}=0 \, , \label{curlK1}\\
K_{ab[cd,i]}=0 \, . \label{curlK2}
\end{eqnarray}

Following Appendix C of \cite{Bunster:2013oaa}, we first decompose the tensor
$K \sim \ydiagram{1,1} \otimes \ydiagram{1,1}$ into three parts of irreducible Young symmetry,
\begin{eqnarray}
K_{abcd}&=&R_{abcd} + (Q_{abcd} - Q_{abdc}) + A_{abcd}\, ,
\end{eqnarray}
where
\begin{eqnarray}
R_{abcd}&=&\frac{1}{2}(K_{abcd} + K_{cdab}) \sim \ydiagram{2,2} \,, \\
Q_{abcd} &=& \frac{3}{2} K_{[abc]d} \sim  \ydiagram{2,1,1} \,, \\
A_{abcd}&=& K_{[abcd]}=0 \sim  \ydiagram{1,1,1,1} \, .
\end{eqnarray}
This permits us to use generalized Poincaré lemmas on each of the irreducible parts.
We start  with the proof
that $\pd_{[i}R_{ab]cd}=0$,
which follows by using \eqref{curlK1} and \eqref{curlK2}
\begin{eqnarray}
\pd_{[i}R_{ab]cd}&=&\frac{1}{2}(\pd_{[i}K_{ab]cd} + K_{cd[ab,i]}) =0 \, . \label{dR=0}
\end{eqnarray}
Using the generalized Poincaré lemma of~\cite{DuboisViolette:1999rd,DuboisViolette:2001jk},
we can write $R_{abcd}$ as 
\begin{eqnarray}
R_{abcd}&=& \pd_{[a}S_{b][c,d]} \sim \ytableaushort{{}{},\pd\pd}\, , 
\end{eqnarray}
where $S_{ab} \sim  \ydiagram{2}$ is a symmetric tensor.
For $Q_{abcd}$, the equation $\pd_{[i}Q_{abc]d}=0$ follows directly from \eqref{curlK1}. To prove that also $Q_{abc[d,i]}=0$,
let us use the fact that the curl of $K$ in the second group of indices vanishes,
\begin{equation}
0=K_{ab[cd,i]} = \frac{1}{3}(K_{abcd,i}+ K_{abdi,c}+K_{abic,d}) = \frac{1}{3}(2 K_{abc[d,i]}+K_{abdi,c}) .
\end{equation}
Then we have that $K_{abc[d,i]}=-\frac{1}{2}K_{abdi,c}$. Finally,
\begin{eqnarray}
Q_{abc[d,i]}&=& \frac{3}{2} K_{[abc][d,i]}\nonumber \\
            &=& \frac{1}{2}(K_{abc[d,i]}+K_{bca[d,i]}+K_{cab[d,i]}) \nonumber \\
            &=& -\frac{1}{4}(\pd_{c} K_{abdi}+\pd_{a}K_{bcdi}+\pd_{b}K_{cadi}) \nonumber \\
            &=& -\frac{3}{4}\pd_{[a}K_{bc]di}=0\, .
\end{eqnarray}
which vanishes provided (\ref{curlK1}). Then, using the generalized
Poincaré lemma of \cite{Bekaert:2002dt} we have 
\begin{eqnarray}
Q_{abcd}=\frac{3}{4}\pd_{d}\pd_{[a}A_{bc]} \sim \ytableaushort{{}\pd,{},\pd}\, ,
\end{eqnarray}
where $A_{ab}\sim \ydiagram{1,1}$ is an antisymmetric tensor. In the definition of $K_{abcd}$ we have the combination
\begin{eqnarray}
Q_{abcd}-Q_{abdc}&=& \frac{3}{4}\pd_{d}\pd_{[a}A_{bc]}-\frac{3}{4}\pd_{c}\pd_{[a}A_{bd]}\nonumber \\
                 &=& \pd_{[a}A_{b][c,d]} \, .
\end{eqnarray}
Finally, the explicit form of $K_{abcd}$ is given by
\begin{eqnarray}
K_{abcd}&=&R_{abcd} + (Q_{abcd} - Q_{abde})    \nonumber\\
        &=& \pd_{[a}S_{b][c,d]} +\pd_{[a}A_{b][c,d]} \nonumber \\
        &=& \pd_{[a}M_{b][c,d]}
\end{eqnarray}
where $M_{ab} = S_{ab}+ A_{ab}$. This finishes the proof.

\section{Conformal geometry of a \texorpdfstring{$(2,2,1)$}{(2,2,1)} field}
\label{sec:conf}

In this appendix, we construct the various curvature tensors of the prepotential $Z_{ijklm}$ associated with the gauge and Weyl transformations \eqref{eq:gaugeZ} -- \eqref{eq:WeylZ}, following the pattern of \cite{Bunster:2012km,Bunster:2013oaa,Henneaux:2015cda,Henneaux:2016zlu,Henneaux:2016opm,Henneaux:2017xsb,Lekeu:2018kul}.

We then prove the two key properties of the Cotton tensor of $Z$:
\begin{enumerate}
\item It provides a complete set of gauge and Weyl invariants, i.e., any gauge and Weyl invariant function of $Z$ can be written as a function of the Cotton tensor and its derivatives only.
\item Any tensor that satisfies the differential, symmetry and trace properties of the Cotton tensor is really the Cotton tensor of some field.
\end{enumerate}
The first of those properties is equivalent to the property
\begin{equation}
D[Z] = 0 \; \Leftrightarrow \; Z = \text{(gauge)} + \text{(Weyl)},
\end{equation}
where by the right-hand side we mean terms of the form \eqref{eq:gaugeZ} -- \eqref{eq:WeylZ}. The second property was dubbed ``conformal Poincaré lemma'' in \cite{Lekeu:2018kul}, since it is a generalization of the Poincaré lemmas of \cite{Olver_hyper,DuboisViolette:1999rd,DuboisViolette:2001jk,Bekaert:2002dt} with the addition of trace conditions and Weyl transformations. The proofs closely follow those of \cite{Henneaux:2015cda,Lekeu:2018kul} for two columns Young tableaux. It is nevertheless useful to spell out the details.

\subsection{Einstein, Schouten and Cotton tensors}
\label{sec:prepot}

The Einstein tensor of $Z\indices{^{pqrjk}}$ is defined as
\begin{equation}
    G\indices{_{de}^{l}}[Z] \equiv \frac{2}{3!4!}\varepsilon\indices{^{l}_{spqr}}\varepsilon_{deijk}\pd^{s}\pd^{i}Z^{pqrjk}\, .
\end{equation}
It has the $(2,1)$ Young symmetry and is identically divergenceless on both groups of indices. It is invariant under the gauge transformations \eqref{eq:gaugeZ}, parametrized by $\alpha$ and $\beta$. The converse of those properties is also true \cite{Bekaert:2002dt}:
\begin{enumerate}
    \item Any divergenceless $(2,1)$ tensor can be written as the Einstein tensor of some $Z$.
    \item The vanishing of the Einstein tensor implies that $Z$ is pure gauge.
\end{enumerate}
The Einstein tensor is not invariant under the Weyl transformations \eqref{eq:WeylZ}. Instead, the tensor that controls Weyl invariance is the Cotton tensor, defined as 
\begin{equation}\label{eq:defD}
    D_{abcde}[Z] \equiv \frac{1}{2}\varepsilon_{abclm}\pd^{m} S\indices{_{de}^{l} }[Z]\, ,
\end{equation}
where the Schouten tensor $S\indices{_{de}^{l}}[Z]$ is defined in terms of the Einstein tensor and its trace as 
\begin{equation}\label{eq:SofG}
S\indices{_{de}^{l}}[Z] \equiv G\indices{_{de}^{l}}[Z] + \frac{2}{3} \delta^{l}_{[d}G\indices{_{e]p}^p}[Z] \, . 
\end{equation}
The Schouten tensor transforms under Weyl rescalings as
\begin{equation}
\delta S\indices{_{de}^{l}}=\frac{1}{18}\pd_{[d}\pd^{l}\rho_{e]}\, . \end{equation}
The relation between the Einstein and the Schouten can be inverted to $G\indices{_{de}^{l}}=S\indices{_{de}^{l}}+2 \delta_{[d}^{l}S\indices{_{e]m}^{m}}$.
It follows from this formula that the divergencelessness of $G\indices{_{de}^{l}}$ is equivalent to the differential identity
\begin{equation}\label{eq:diffidS}
    \pd^{j}S_{ijk}-\pd_{k}S\indices{_{ij}^{j}} = 0
\end{equation}
on the Schouten and its trace.

The Cotton tensor $D_{abcde}$ is of irreducible Young symmetry type $(2,2,1)$, identically transverse in both group of indices, $\pd_{a}D\indices{^{abc}_{de}}=0=\pd^{d}D\indices{^{abc}_{de}}$, and double-traceless, $D\indices{^{abc}_{bc}}=0$. This follows from the identity \eqref{eq:diffidS}, the definition \eqref{eq:defD} (and the previous property for the second group of indices), and the cyclic identity $S_{[del]} = 0$, respectively. Additionally, the Cotton tensor is invariant under the transformations \eqref{eq:gaugeZ} -- \eqref{eq:WeylZ}. The converse of those two properties are proved below.

\subsection{Gauge completeness}
\label{app:th1}

Since the converse is true by construction (see section \ref{sec:prepot}), we will only prove the implication $D[Z] = 0 \; \Rightarrow \; Z = \text{(gauge)} + \text{(Weyl)}$.

First,
\begin{equation}
    D_{abcde}=\frac{1}{2}\varepsilon_{abclm}\pd^{m}S\indices{_{de}^{l}}=0
\end{equation}
implies that $S\indices{_{de}^{[l,m]}}=0$. This property, along with the cyclic identity $S_{[dem]} = 0$, also implies that $\pd_{[m}S\indices{_{de]}^{l}}=0$ :
\begin{align}
\pd_{[m}S\indices{_{de]l}} &= \frac{1}{3}\left(\pd_{m} S\indices{_{del}} +\pd_{d} S\indices{_{eml}}+\pd_{e} S\indices{_{mdl}} \right) = \frac{1}{3}\left(\pd_{l} S\indices{_{dem}} +\pd_{l} S\indices{_{emd}}+\pd_{l} S\indices{_{mde}} \right) = \pd_{l}S_{[dem]} \nonumber \\
                 &= 0  \, .
\end{align}
Since its curl on both groups of indices vanishes, the generalized Poincaré lemma of \cite{Bekaert:2002dt} implies that the tensor $S\indices{_{de}^{m}}$ can be written as
\begin{equation}
    S\indices{_{de}^{m}}=\frac{1}{18}\pd_{[d}\pd^{m} \rho_{e]}\, ,
\end{equation}
for some $\rho$.
This implies that the Einstein tensor of $Z$ is equal to
\begin{equation}
    G\indices{_{de}^{l}} = \frac{2}{3!4!}\varepsilon\indices{^{l}_{spqr}}\varepsilon_{deijk}\pd^{s}\pd^{i}\delta^{pj}\delta^{qk}\rho^r\, .
\end{equation}
Therefore, the Einstein tensor of $Z\indices{^{pqr}_{jk}} - \delta^{[p}_{[j}\delta^{q}_{k]}\rho^{r]}$ is zero, i.e., this quantity is pure gauge. This shows the result.

\subsection{Conformal Poincaré lemma}
\label{app:th2}

This appendix is devoted to the proof that, for any tensor $T$ that satisfies the three properties
\begin{enumerate}
    \item It is of the type $(2,2,1)$,
    \item It is transverse in both groups of indices, 
    \item It is double-traceless,
\end{enumerate}
there exists a field $Z$ such that $T$ is the Cotton tensor of $Z$, i.e., $T=D[Z]$. The fact that the Cotton tensor satisfies these three properties was proven in section \ref{sec:prepot}. Moreover, $Z$ is determined from $T$ up to the gauge and Weyl transformations of section \ref{sec:chiralaction}, as follows from the first property of the Cotton tensor.

The first step is to introduce a tensor $P_{ijkl}$ such that 
\begin{equation}\label{pofT}
    P_{ijkl}=\frac{1}{3!}T\indices{^{abc}_{ij}}\varepsilon_{klabc}\, .
\end{equation}
The properties listed above are equivalent to
\begin{enumerate}
    \item $P$ is traceless, $P\indices{^i_{jil}} = 0\,$, 
    \item $P_{ij[kl,m]}=0$ and $\pd^i P_{ijkl} =0\,$,
    \item $P_{[ijkl]}=0\,$.
\end{enumerate}
Now, we would like to prove that $P_{ijkl}=S_{ij[k,l]}$ where $S$ has the $(2,1)$ symmetry, $S  \sim {\tiny \ydiagram{2,1}}$. In a second step, $S$ will be identified with the Schouten tensor of some $Z$. Using the usual Poincar\'{e} lemma, $P_{ij[kl,m]}=0$ implies
\begin{equation}\label{eq:PofM}
P_{ijkl}= M_{ij[k,l]}\, ,
\end{equation}
where the tensor $M_{ijk}=M_{[ij]k}$ is not \textit{a priori} of irreducible Young symmetry. In order to have $(2,1)$ Young symmetry for $M_{ijk}$, i.e., $M_{[ijk]}=0$, we can use the freedom $\tilde{M}_{ijk} = M_{ijk} + \pd_{k} A_{ij}$ in \eqref{eq:PofM}, where $A_{ij}$ is antisymmetric. Now, the fact that $P_{[ijkl]}=0$ implies that $M_{[ijk,l]}=0$, from which we get, using the usual Poincar\'{e} lemma again, 
\begin{equation}
M_{[ijk]} = \pd_{[i}\lambda_{jk]} \, .
\end{equation}
However, we can fix the ambiguity above as $A_{ij}=-\lambda_{ij}$ to have $\tilde{M}_{[ijk]}=0$. Thus, $\tilde{M}_{ijk} \equiv S_{ijk}$ has Young symmetry $(2,1)$.

Now, we would like to prove that there exists a $Z$ such that $S = S[Z]$. For that, we use the fact that $P$ is traceless,
\begin{eqnarray}
    0&=& 2 \delta^{jl}P_{ijkl} \nonumber \\
     &=& \pd^{j}S_{ijk}-\pd_{k}S\indices{_{ij}^{j}}\, ,
\end{eqnarray}
which is equivalent to the divergencelessness of the tensor $G$ defined in terms of $S$ by inverting relation \eqref{eq:SofG}.
Now, it follows from the generalized Poincaré lemma of \cite{Bekaert:2002dt} that for any divergenceless tensor $G$ of $(2,1)$ Young symmetry, there exists a $Z$ such that $G=G[Z]$. The relation between the Schouten and Einstein tensors gives then $S=S[Z]$.
Finally, we can conclude that $T=D[Z]$ by plugging back into \eqref{pofT}:
\begin{equation}
    T\indices{^{abc}_{ij}} = \frac{1}{2}P_{ijkl}\varepsilon^{klabc}
     = \frac{1}{2} S_{ijk,l}[Z]\varepsilon^{klabc} = D\indices{^{abc}_{ij}}[Z]\, .
\end{equation}

\section{Split of the non-chiral action}
\label{app:ham}

In this appendix, we show that the action \eqref{actionZ} for the chiral $(2,1)$-tensor can also be obtained by splitting the first-order (Hamiltonian) action for a non-chiral tensor into its chiral and anti-chiral parts. We follow references \cite{Deser:1997se,Bekaert:1998yp}, where it was done for the case of chiral two-forms in six dimensions (see also Appendix D of \cite{Henneaux:2016opm} for the case of the chiral $(2,2)$-tensor).

\subsection{Hamiltonian and constraints}

The Lagrangian for a non-chiral $(2,1)$-tensor $\phi_{\mu\nu\rho}$ is given by \cite{Curtright:1980yk}
\begin{equation}
\mathcal{L} = - 12\, \delta^{\mu_1 \mu_2 \mu_3 \mu_4}_{\nu_1 \nu_2 \nu_3 \nu_4} M\indices{^{\nu_1 \nu_2 \nu_3}_{\mu_4}} M\indices{_{\mu_1 \mu_2 \mu_3}^{\nu_4}}\, ,
\end{equation}
with $M_{\mu\nu\rho\sigma} = \partial_{[\mu} \phi_{\nu\rho]\sigma}$.
The associated Hamiltonian action is
\begin{equation}
S_H = \int \! dt \, d^5\! x \, \left( \pi_{ijk} \dot{\phi}^{ijk} - \mathcal{H} - n_{ij} \,\mathcal{C}^{ij} - N_{i}\, \mathcal{C}^{i} \right) \, ,
\end{equation}
where the Hamiltonian density is
\begin{align}
\mathcal{H} &= \mathcal{H}_\pi + \mathcal{H}_\phi \, , \\
\mathcal{H}_\pi &= \frac{1}{4} \left( \pi^{ijk}\pi_{ijk} + \frac{2}{3} \pi\indices{^{ij}_j} \pi\indices{_{ik}^k} \right) \, , \\
\mathcal{H}_\phi &= 12\, \delta^{i_1 \dots i_4}_{j_1 \dots j_4} \, M\indices{^{j_1 j_2 j_3}_{i_4}} \, M\indices{_{i_1 i_2 i_3}^{j_4}} \, .
\end{align}
The components $n_{ij} = \phi_{0ij}$ and $N_{i} = \phi_{i00}$ of $\phi_{\mu\nu\rho}$ with some indices equal to zero only appear as Lagrange multipliers for the constraints
\begin{align}
\mathcal{C}^{ij} &\equiv 2 \,\partial_k \left( \pi^{ijk} - \pi^{kij} \right) = 0\, ,\\
\mathcal{C}_i &\equiv 12 \,\delta^{abc}_{ijk} \, \partial^j \partial_a \phi\indices{_{bc}^k} = 0 \, .
\end{align}
The constraint $\mathcal{C}^{ij} = 0$ is equivalent to
\begin{equation}\label{mom2}
\partial_k \pi^{kij} = 0 \qquad ( \Rightarrow \; \partial_k \pi^{ijk} = 0)
\end{equation}
because of the cyclic identity $\pi^{[ijk]} = 0$. The constraint $\mathcal{C}_i = 0$ is equivalent to the double tracelessness of the electric field defined in section \ref{sec:electr-magn-field},
\begin{equation}
\mathcal{E}\indices{_{ijk}^{jk}}[\phi] = 0 \, .
\end{equation}

\subsection{Prepotentials}

The momentum constraint is solved by introducing a first prepotential $Z^{(1)}_{ijk lm}$ of $(2,2,1)$ symmetry, in terms of which the momentum reads
\begin{equation}
\pi_{ijk} = 6 \, G_{ijk}[Z^{(1)}]
\end{equation}
(we include an extra factor for convenience). This follows from the properties of the Einstein tensor defined in Appendix \ref{sec:prepot}. The Hamiltonian constraint is solved by introducing a second prepotential $Z^{(2)}_{ijk lm}$, as in formula \eqref{phiofZ}. The various terms in the Hamiltonian action are then, up to a total derivative,
\begin{align}
\pi_{ijk} \dot{\phi}^{ijk} &= 2\, Z^{(1)}_{abcij} \dot{D}^{abcij}[Z^{(2)}] \, , \\
\mathcal{H}_\pi &= 18\, G_{ijk}[Z^{(1)}] S^{ijk}[Z^{(1)}] = \frac{1}{2} Z^{(1)\, abc pq} \varepsilon_{pqijk} \partial^{k} D\indices{_{abc}^{ij}}[Z^{(1)}]\, , \\
\mathcal{H}_\phi &= 18\, G_{ijk}[Z^{(2)}] S^{ijk}[Z^{(2)}] = \frac{1}{2} Z^{(2) abc pq} \varepsilon_{pqijk} \partial^{k} D\indices{_{abc}^{ij}}[Z^{(2)}]\, .
\end{align}
Since the constraints are identically satisfied, the Lagrange multipliers $n_{ij}$ and $N_i$ disappear from the action.

We now do the change of variables
\begin{equation}
Z^\pm = Z^{(1)} \pm Z^{(2)} \; \Leftrightarrow \; Z^{(1)} = \frac{1}{2} \left( Z^+ + Z^- \right)\, , \; Z^{(2)} = \frac{1}{2} \left( Z^+ - Z^- \right)\, .
\end{equation}
This splits the action into two parts, $S[Z^+, Z^-] = S^+[Z^+] - S^-[Z^-]$, where $S^+$ is exactly the action \eqref{actionZ} obtained in section \ref{sec:chiralaction}. The action $S^{-}$ is the same, but with the sign of the second term flipped: it is the action for an anti-chiral $(2,1)$-tensor.

\section{Dimensional reduction}
\label{sec:dimred}

In this appendix, we perform the dimensional reduction of the chiral $(2,1)$-tensor. At the level of the field $\phi_{\mu\nu\rho}$ itself, we get upon dimensional reduction the four fields
\begin{equation}
\phi_{\mu\nu\rho}, \quad \phi_{5(\mu\nu)}, \quad \phi_{5[\mu\nu]}, \quad \phi_{\mu55} .
\end{equation}
The interpretation of those fields is as follows: the symmetric tensor $\phi_{5(\mu\nu)}$ is the five-dimensional graviton, the mixed symmetry field $\phi_{\mu\nu\rho}$ is the dual graviton (Curtright field), $\phi_{\mu55}$ is a vector field, and the two-form $\phi_{5[\mu\nu]}$ is the magnetic dual of $\phi_{\mu55}$. The fact that $\phi_{\mu\nu\rho}$ is not independent from $\phi_{5(\mu\nu)}$, and that $\phi_{5[\mu\nu]}$ is not independent from $\phi_{\nu55}$, follows from the self-duality condition in $D=6$ \cite{Hull:2000zn}. All in all, this describes one metric and one vector field.

We now show that the reduction of the prepotential $Z\indices{^{IJK}_{AB}}$ correctly reproduces the prepotentials of a metric and a vector field in five dimensions, and that the actions also match (see \cite{Bunster:2013oaa} for five-dimensional linearized gravity and \cite{Bunster:2011qp} for the two-potential formulation of the free vector field). In this appendix, capital indices go from one to five, and lowercase indices only go from one to four.

The prepotential $Z\indices{^{IJK}_{AB}}$ splits into
\begin{equation}
Z\indices{^{ijk}_{ab}}\, , \quad Z\indices{^{ij5}_{ab}}\, , \quad Z\indices{^{ijk}_{a5}}\, , \quad Z\indices{^{ij5}_{a5}}\, .
\end{equation}
All the pieces have irreducible Young symmetry except for $Z\indices{^{ij5}_{ab}}$, which has components of {\tiny$\ydiagram{2,2}$} and {\tiny$\ydiagram{2,1,1}$} symmetry.
The cyclic identity $Z_{[IJKL]M} = 0$ implies
\begin{equation}
Z_{5[ijk]l} = - \frac{1}{3} Z_{ijkl5}\, ,
\end{equation}
which gives $Z_{5[ijkl]} = 0$. We can then split $Z\indices{^{ij5}_{ab}}$ in irreducible parts as
\begin{align}
Z\indices{^{ij5}_{ab}} &= R\indices{^{ij}_{ab}} + 2 Q\indices{^{ij}_{[ab]}}\, ,
\end{align}
with
\begin{align}
R_{ijab} &= \mathbb{P}_{(2,2)} \left( Z_{ij5ab} \right) = \frac{1}{2} (Z_{ij5ab} + Z_{ab5ij})\, , \\
Q_{ijab} &= \mathbb{P}_{(2,1,1)} \left( Z_{ij5ab} \right) = \frac{3}{2} Z_{5[ija]b} = - \frac{1}{2} Z_{ijab5}\, .
\end{align}
For the other two irreducible components, we make the triangular change of variables
\begin{align}
M\indices{^{ij}_a} &= Z\indices{^{ij5}_{a5}} \,, \\
N\indices{^{ijk}_{ab}} &= Z\indices{^{ijk}_{ab}} + 3 \delta^{[i}_{[a} Z\indices{^{jk]5}_{b]5}} \, .
\end{align}
The tensors $M$ and $N$ have $(2,1)$ and $(2,1,1)$ Young symmetries, respectively. The following inversion formulas are useful for explicit computations:
\begin{align}
Z\indices{^{ijk}_{ab}} &= N\indices{^{ijk}_{ab}} - 3 \delta^{[i}_{[a} M\indices{^{jk]}_{b]}}\, , \\
Z\indices{^{ij5}_{ab}} &= R\indices{^{ij}_{ab}} + 2 Q\indices{^{ij}_{[ab]}}\, , \\
Z_{ijab5} &= - 2 Q_{ijab}\, , \\
Z\indices{^{ij5}_{a5}} &= M\indices{^{ij}_a}\, .
\end{align}
With these definitions, the spatial components $\phi_{IJK}$ of the chiral tensor decompose (according to \eqref{phiofZ}) as
\begin{align}
\phi_{ijk} &= \frac{1}{6} \pd^a \left( R\indices{^{bc}_{ij}} \varepsilon_{kabc} - R\indices{^{bc}_{k[i}} \varepsilon_{j]abc} \right)\, , \label{eq:phiofR}\\
\phi_{5(ij)} &= - \frac{1}{4} \partial^a M\indices{^{bc}_{(i}} \varepsilon_{j)abc}\, , \label{eq:phiofM}\\
\phi_{5[ij]} &= - \frac{1}{36} \partial^a N\indices{^{bcd}_{ij}}\varepsilon_{abcd}\, , \\
\phi_{i55} &= - \frac{1}{6} \pd^a Q\indices{^{bcd}_i} \varepsilon_{abcd}\, ,
\end{align}
up to a gauge transformation (the fact that the contribution of $Q$ to $\phi_{ijk}$ is pure gauge was proved in \cite{Bunster:2013oaa}).
Up to constant factors, equations \eqref{eq:phiofM} and \eqref{eq:phiofR} are exactly the expressions given in reference \cite{Bunster:2013oaa} for the metric and its dual in terms of the prepotentials for five-dimensional gravity. Moreover, the last two equations motivate the definitions
\begin{equation}
V_i = \pd^a Q\indices{^{bcd}_i} \varepsilon_{abcd}, \quad W_{ij} = \partial^a N\indices{^{bcd}_{ij}}\varepsilon_{abcd},
\end{equation}
for the two potentials of the vector fields in five dimensions (up to factors that will be fixed below).

The gauge and Weyl transformations of the lower-dimensional prepotentials also match: the gauge and Weyl transformations of $R_{ijab}$ and $M_{ija}$ are exactly those of the prepotentials of five-dimensional linearized gravity. The prepotentials $N_{ijkab}$ and $Q_{abci}$ have no Weyl transformation, and their gauge transformations are exactly such that $V_i$ and $W_{ij}$ transform as total derivatives.

Now, the Cotton tensor of $Z$ reduces as follows:
\begin{align}
D_{abcde}[Z] &= \frac{1}{3!^{2}} \varepsilon_{abcj} \pd^j \cB_{de}[V]\, , \label{Dabcde}\\
D_{abcd5}[Z] &= \frac{1}{3 \cdot 3!^2} \varepsilon_{abcj} \partial^j \cB_d[W]\, ,\label{Dabcd5} \\
D_{ab5de}[Z] &= -\frac{1}{4 \cdot 3!} D_{abde}[M] + \frac{1}{3 \cdot 3!^2} \varepsilon_{abj[d}\pd^j \cB_{e]}[W]\, ,\label{Dab5de} \\
D_{ab5d5}[Z] &= \frac{1}{2 \cdot 4!}D_{abd}[R] - \frac{1}{2 \cdot 3!^{2}} \varepsilon_{abij}\pd^i \cB\indices{^j_d}[V]\, .\label{Dab5d5}
\end{align}
Here, the magnetic fields are \cite{Bunster:2011qp}
\begin{equation}
\cB_{ij}[V] = \varepsilon_{ijkl}\pd^k V^l, \quad \cB_{i}[W] = \frac{1}{2} \varepsilon_{ijkl} \pd^j W^{kl} .
\end{equation}
The definitions and properties of the Cotton tensors of the gravity prepotentials can be found in \cite{Lekeu:2018kul}.
Using these formulas, the reduction of the action \eqref{actionZ} is direct and gives the sum of the actions \eqref{eq:lingrav} and \eqref{eq:vector} below, up to some factors that are fixed in \eqref{eq:factors}.

The action of five-dimensional gravity is given in the prepotential formalism by
\begin{align}\label{eq:lingrav}
S[\Phi, P] = \int\!dt\, d^4\!x\, \big( &D_{ijkl}[\Phi] \, \dot{P}^{ijkl} - D_{ijk}[P] \,\dot{\Phi}^{ijk} \\
&- P_{ijkl}\, \varepsilon^{ij pq} \pd_p D\indices{^{kl}_q}[P] - \frac{1}{2} \Phi_{ijk} \, \varepsilon^{abkl} \pd_l D\indices{^{ij}_{ab}}[\Phi] \big) \, , \nonumber
\end{align}
where $\Phi_{ijk}$ and $P_{ijkl}$ are the prepotentials with $(2,1)$ and $(2,2)$ symmetry. This action was first written in \cite{Bunster:2013oaa}; see \cite{Lekeu:2018kul} for the rewriting in terms of the appropriate Cotton tensors defined there.
The action of a five-dimensional vector field is \cite{Bunster:2011qp}
\begin{align}\label{eq:vector}
S[v,w] = \frac{1}{2} \int\!dt\, d^4\!x\, \big( &\cB_i[w] \dot{v}^i - \frac{1}{2} \cB_{ij}[v] \dot{w}^{ij} \\
&- \frac{1}{2} w_{ij} \varepsilon^{ijkl} \pd_k \cB_l[w] - \frac{1}{2} v_i \varepsilon^{ijkl} \partial_j \cB_{kl}[v] \big) \, ,\nonumber
\end{align}
where $v_i$ is a vector and $w_{ij}$ a two-form.

Comparing with the reduction of \eqref{actionZ}, we identify the prepotentials as
\begin{equation}\label{eq:factors}
R_{ijkl} = 4 \sqrt{2} \, P_{ijkl}\,,\quad M_{ijk} = 2 \sqrt{2} \, \Phi_{ijk}\,,\quad V_i = \sqrt{6}\, v_i\,,\quad W_{ij} = 3\sqrt{6}\, w_{ij}\,. 
\end{equation}

\section{Maximal supergravity in five dimensions}
\label{app:susy}

In this appendix, we write the action and supersymmetry transformations of linearized maximal supergravity in five dimensions \cite{Cremmer:1979uq} in the first-order (prepotential) formalism. It can be obtained by dimensional reduction of the $\cN = (4,0)$ theory of \cite{Henneaux:2017xsb}, or by reduction of the $\cN = (3,1)$ theory presented in this paper. Comparison of the two enables us to fix the $\beta_1, \dots, \beta_{10}$ coefficients that were left undetermined in section \ref{sec:susy}. Capital indices $A,B,\dots$ are $\mathfrak{usp}(8)$ indices (running from $1$ to $8$).

\subsection{Field content and linearized first-order action}
\label{sec:5daction}

\paragraph{Field content.} The field content of maximal supergravity in five dimensions is the following \cite{Cremmer:1979uq}: one metric $g_{\mu\nu}$, $8$ gravitinos $\Psi_\mu^A$, $27$ vectors $V^{AB}_\mu$, $48$ spin $1/2$ fields $\Psi^{ABC}$, and $42$ scalar fields $\Phi^{ABCD}$. They satisfy the appropriate $\mathfrak{usp}(8)$ irreducibility and reality conditions (there is no chirality in five dimensions). For fermions, those involve the five-dimensional analogue of the $\mcB$-matrix, defined in this case by $\gamma_\mu^* = - \mcB_{(5)} \gamma_\mu \mcB_{(5)}^{-1}$.

\paragraph{Linearized action.} The first-order action for the linearized theory is the sum of the following five terms:
\begin{itemize}
\item The metric is described by two real prepotentials
\begin{equation}
\phi_{ijk} \sim {\tiny\ydiagram{2,1}}\,, \quad P_{ijkl} \sim {\tiny\ydiagram{2,2}}\,,
\end{equation}
with action
\begin{align}
S_2[\phi, P] = \int\!dt\, d^4\!x\, \big( &D_{ijkl}[\phi] \, \dot{P}^{ijkl} - D_{ijk}[P] \,\dot{\phi}^{ijk} \\
&- P_{ijkl}\, \varepsilon^{ij pq} \pd_p D\indices{^{kl}_q}[P] - \frac{1}{2} \phi_{ijk} \, \varepsilon^{abkl} \pd_l D\indices{^{ij}_{ab}}[\phi] \big) \,. \nonumber
\end{align}

\item The $8$ gravitinos are described by the prepotentials
\begin{equation}
\Theta^{A}_{ij}\, ,
\end{equation}
with action
\begin{equation}
S_{\frac{3}{2}}[\Theta] = - i \int \!dt \,d^4\!x\, \Theta^\dagger_{Aij} \left( \dot{D}^{Aij}[\Theta] - i \varepsilon^{ijkl} \gamma^m \partial_k D^A_{lm}[\Theta] \right) \,.
\end{equation}
This action, and the definition of the Cotton tensor $D[\Theta]$, can be found in \cite{Lekeu:2018kul}.

\item The $27$ vectors are described by the two potentials
\begin{equation}
V_i^{AB}\,, \quad W_{ij}^{AB}\,.
\end{equation}
The action is \cite{Bunster:2011qp}
\begin{align}
S_1[V,W] = \frac{1}{2} \int\!dt\, d^4\!x\, \big( &\dot{V}^{*i}_{AB} \cB^{AB}_i[W] - \frac{1}{2} \dot{W}^{*ij}_{AB}\cB^{AB}_{ij}[V] \\
&- \frac{1}{2} W^*_{ABij} \varepsilon^{ijkl} \pd_k \cB^{AB}_l[W] - \frac{1}{2} V^*_{ABi} \varepsilon^{ijkl} \partial_j \cB^{AB}_{kl}[V] \big)\,.\nonumber
\end{align}

\item The action for the 48 spin $1/2$ fields is
\begin{equation}
S_{\frac{1}{2}}[\Psi] = i \int \!dt \,d^4\!x\, \Psi^{\dagger}_{ABC} \left(\dot{\Psi}^{ABC} - \gamma^{0} \gamma^{i} \pd_{i} \Psi^{ABC} \right) \,.
\end{equation}

\item The 42 scalars are described by the usual Hamiltonian action
\begin{equation}
    S_0[\Phi,\Pi] = \frac{1}{2} \int \!dt \,d^4\!x\, \left(2\, \Pi^{*}_{ABCD} \dot{\Phi}^{ABCD} - \Pi_{ABCD}^{*} \Pi^{ABCD} - \pd_{i} \Phi_{ABCD}^{*} \pd^{i} \Phi^{ABCD} \right)\, .
\end{equation}

\end{itemize}

\subsection{Dimensional reduction of the \texorpdfstring{$\cN = (4,0)$}{N = (4,0)} theory}
\label{sec:SUSY40}

\paragraph{Reduction of the action.} Five-dimensional maximal supergravity in the prepotential formalism, as described above, can be obtained from direct dimensional reduction of the $\cN = (4,0)$ theory in six dimensions \cite{Henneaux:2017xsb} using the following identifications (we write the higher-dimensional quantities with a hat):
\begin{itemize}
\item The prepotential $\hat{Z}_{IJKL}$ for the exotic graviton splits as
\begin{equation}
\hat{Z}_{ijkl} = 12 \sqrt{3}\, P_{ijkl}\,, \quad \hat{Z}_{ijk5} = 3\sqrt{3}\, \phi_{ijk}\,, \quad \hat{Z}_{i5j5}=0.
\end{equation}
\item The chiral $2$-forms $\hat{A}^{AB}_{IJ}$ give
\begin{equation}
\hat{A}^{AB}_{i5} = V_i^{AB} / \sqrt{2}\,, \quad \hat{A}^{AB}_{ij} = - W_{ij}^{AB} / \sqrt{2}\,.
\end{equation}
\item The chiral fermionic prepotential $\hat{\chi}^A_{IJ}$ is
\begin{equation}
\hat{\chi}^{A}_{IJ} = \begin{pmatrix} \hat{\chi}^{A+}_{IJ} \\ 0 \end{pmatrix}\,, \quad \hat{\chi}^{A+}_{ij} = \Theta^{A}_{ij} / \sqrt{2}  \,, \quad \hat{\chi}^{A+}_{i5}=0\,.
\end{equation}
\item The scalars stay the same, and the Dirac fields are simply
\begin{equation}
\hat{\Psi}^{ABC} = \begin{pmatrix} \Psi^{ABC} \\ 0 \end{pmatrix}\, .
\end{equation}
\end{itemize}
This information is gathered in \cite{Lekeu:2018kul}. The fermions have component only in the first half; this is due to the chirality conditions in six dimensions. The form of the gamma matrices we use in the reduction can be found in section 6 of \cite{Henneaux:2017xsb}. Their form implies that the $\mcB$ matrix appearing in the reality conditions in six dimensions takes the form
\begin{equation}\label{eq:realitymatrix65}
    \mcB_{(6)} = \begin{pmatrix} \mcB_{(5)} & 0 \\ 0 & -\mcB_{(5)} \end{pmatrix}\, ,
\end{equation}
where $\mcB_{(5)}$ is the analog matrix in five dimensions. Therefore, the reality conditions in six and five dimensions agree.

\paragraph{Reduction of the supersymmetry transformations.} We can also find the supersymmetry transformations in five dimensions from the reduction of those of the $(4,0)$-theory. In this reduction, the supersymmetry parameter of the $(4,0)$ theory reduces as\footnote{The factor of $i$ comes from the form \eqref{eq:realitymatrix65} of the reality matrix: it is such that the reality condition $\hat{\epsilon}^*_A = \Omega_{AB} \mcB_{(6)} \hat{\epsilon}^B$ in six dimensions implies correctly $\epsilon^*_A = \Omega_{AB} \mcB_{(5)} \epsilon^B$ in five.}
\begin{equation}\label{eq:SUSYparam40}
\hat{\epsilon}^{A} = \begin{pmatrix} 0 \\ i \epsilon^{A} \end{pmatrix} \,.
\end{equation}
With these identifications, we get the five-dimensional supersymmetry transformations\footnote{As was stressed in \cite{Lekeu:2018kul}, the supersymmetry transformation of the gravitino prepotentials $\Theta^{A}_{ij}$ picks up an extra term when enforcing the gauge condition $\Theta^{A}_{i5} = 0$.}
\begin{align}
\delta P_{ijkl} &= - \frac{i \alpha_1}{12 \sqrt{6}} \,{\mathbb P}_{(2,2)} \left( \bar{\epsilon}_{A} \gamma_{ij} \Theta_{kl}^{A} \right) \\
\delta \phi_{ijk} &= \frac{\alpha_1}{6\sqrt{6}}  \, {\mathbb P}_{(2,1)}\left(\bar{\epsilon}_{A} \gamma_{k} \Theta_{ij}^{A}\right) \\
\delta \Theta^{A}_{ij} &= - \frac{\alpha_1}{12\sqrt{6}} \left( 2 \varepsilon_{qrkl}\pd^{r}P\indices{_{ij}^{kl}} \gamma^{q} + i \varepsilon_{pqrk}\pd^{r} \phi\indices{_{ij}^{k}}\gamma^{pq} - i \pd^r \phi\indices{^{kl}_{[i}} \varepsilon_{j]rkl} \right) \gamma^{0} \epsilon^{A} \nonumber \\
&\quad - \alpha_2 \left( \frac{i}{2} W^{AB}_{ij} +   \gamma_{[i}V^{AB}_{j]} \right) \Omega_{BC} \gamma^{0}\epsilon^{C} \\
\delta W^{AB}_{ij} &= -\alpha_2 \left(2  \bar{\epsilon}_{C} \gamma_{[i} S^{[A}_{j]}[\Theta]\Omega^{B]C} + \frac{1}{4} \Omega^{AB} \bar{\epsilon}_{C} \gamma_{[i}S^{C}_{j]}[\Theta]\right)+ \alpha_{3} \left(i\sqrt{2} \bar{\epsilon}_{C}\gamma_{ij} \Psi^{ABC}\right) \\
\delta V^{AB}_{i} &= -\alpha_2i \left( 2 \bar{\epsilon}_{C} S^{[A}_{i}[\Theta] \Omega^{B]C}+ \frac{1}{4} \Omega^{AB} \bar{\epsilon}_{C} S^{C}_{i}[\Theta]\right) + \alpha_3 \sqrt{2} \left( \bar{\epsilon}_{C} \gamma_{i} \Psi^{ABC} \right) \\
\delta \Psi^{ABC} &= -\alpha_3\frac{1}{2\sqrt{2}} i\gamma^{ij} \gamma^{0} \left(B^{[A B}_{ij}[V]  \epsilon^{C]} - \frac{1}{3}\Omega^{[AB}B^{C]D}_{ij}[V]\Omega_{DE} \epsilon^{E} \right)  \nonumber \\
&\quad +\alpha_3\frac{1}{\sqrt{2}}\gamma^{i}\gamma^{0} \left(B^{[AB}_{i}[W] \epsilon^{C]}-\frac{1}{3} \Omega^{[AB}B^{C]D}_{i}[W] \Omega_{DE}  \epsilon^{E} \right)\nonumber \\
&\quad +\alpha_4 \left(i \Pi^{ABCD}\Omega_{DE}\gamma^{0} \epsilon^{E} + i \pd_{i} \Phi^{ABCD} \Omega_{DE} \gamma^{i} \epsilon^{E}\right) \\
\delta \Phi^{ABCD} &= \alpha_4 \left( -2 i \bar{\epsilon}_{E} \Psi^{[ABC}\Omega^{D]E} -\frac{3}{2}i \bar{\epsilon}_{E} \Omega^{[AB} \Psi^{CD]E}\right) \\
\delta \Pi^{ABCD} &= \alpha_4 \left( -2i \bar{\epsilon}_{E} \gamma^{0} \gamma^{i} \pd_{i} \Psi^{[ABC}\Omega^{D]E} - \frac{3}{2}i \bar{\epsilon}_{E}\gamma^{0} \gamma^{i} \Omega^{[AB}\pd_{i} \Psi^{CD]E}\right) \, .
\end{align}

The constants $\alpha_i$ are determined up to an overall normalization\footnote{As mentioned in \cite{Lekeu:2018kul}, requiring that the supersymmetry variations of the five-dimensional linearized metric and gravitinos have the usual normalization
\[ 
\delta h_{\mu\nu} = \bar{\epsilon}_A\,\gamma_{(\mu} \Psi^A_{\nu)}\, , \quad \delta \Psi^A_{\mu} = \frac{1}{4}\, \partial_{\rho} h_{\mu\nu}\, \gamma^{\nu\rho}\epsilon^A \, ,
\]
fixes $\alpha_1 = 3\sqrt{6}$ (i.e. $\kappa^2 = 1/2$).} by the relations
\begin{equation}\label{eq:relalphas}
\frac{2\alpha_1^2}{(3!)^3} = \alpha_2^2 = \frac{2\alpha_3^2}{3} = \frac{\alpha_4^2}{2} \equiv \kappa^2
\end{equation}
that follow from the supersymmetry algebra \cite{Henneaux:2017xsb}.

\subsection{Dimensional reduction of the \texorpdfstring{$\cN = (3,1)$}{N = (3,1)} theory}

\paragraph{Formulas for dimensional reduction.} The dimensional reduction of the $\cN = (3,1)$ theory in six dimensions goes as follows:
\begin{itemize}
\item The reduction of the prepotential $Z_{IJKLM}$ was done in section \ref{sec:dimred}.

\item The reduction of the chiral two-forms and exotic gravitinos are the same as in the case of the $\cN = (4,0)$ theory, but with the appropriate symplectic indices.

\item From the reduction of $\hat{\theta}^\tta_{IJK}$, we get the prepotential for the gravitinos,
\begin{equation}
\hat{\theta}^\tta_{IJK} = \begin{pmatrix} \hat{\theta}^{\tta +}_{IJK} \\ 0 \end{pmatrix}\, , \quad \theta^\tta_{ij} = 3\hat{\theta}^{\tta +}_{ij5}\, ,
\end{equation}
and also one spin $1/2$ field
\begin{equation}
\psi^\tta = \varepsilon^{ijkl} \partial_i \zeta^\tta_{jkl}\, , \quad \zeta^\tta_{jkl} =  i\sqrt{\frac{3}{4}} \left( i \hat{\theta}^{\tta +}_{jkl} + \gamma_{[j} \hat{\theta}^{\tta +}_{kl]5} \right)\, .
\end{equation}
This reduction was done in the prepotential formalism in \cite{Lekeu:2018kul}; we have added an extra factor of $i$ in $\psi^\tta$ with respect to that reference in order to take the five-dimensional reality condition into account.
\item From the vector field, we have the two potentials
\begin{align}
v^{\tta\ttb}_i = \hat{V}^{\tta\ttb}_i\, , \quad w^{\tta\ttb}_{ij} = \hat{W}^{\tta\ttb}_{ij5}
\end{align}
for vector fields in five dimensions, and also the scalar fields
\begin{align}
\phi^{\tta\ttb} &= \hat{V}^{\tta\ttb}_5 \, , \\
\pi^{\tta\ttb} &= \varepsilon^{abcd} \pd_a \omega^{\tta\ttb}_{bcd}\, , \quad \omega^{\tta\ttb}_{bcd} = \frac{1}{3!} \hat{W}^{\tta\ttb}_{bcd}\, .
\end{align}

\item Because of the chirality conditions, the Dirac fields reduce as
\begin{equation}
\hat{\psi}^{\tta\ttb\alpha} = \begin{pmatrix} \psi^{\tta\ttb\alpha} \\ 0 \end{pmatrix}\, , \quad \hat{\tilde{\psi}}^{\tta\ttb\ttc} = \begin{pmatrix} 0 \\ i \psi^{\tta\ttb\ttc} \end{pmatrix}\, .
\end{equation}

\item The scalar fields stay the same.

\item For the supersymmetry parameters, we have
\begin{equation}
\hat{\epsilon}^{\tta} = \begin{pmatrix} 0 \\ i \epsilon^{\tta} \end{pmatrix}\, , \quad \hat{\tilde{\epsilon}}^{\alpha} = \begin{pmatrix} \epsilon^\alpha \\ 0 \end{pmatrix} \,.
\end{equation} 
\end{itemize}

\paragraph{Index split.} To make contact with the $\mathfrak{usp}(8)$-invariant five-dimensional theory, we split its indices as $A = (\tta\, , \alpha)$, with $\tta=1,\dots,6$ and $\alpha = 1 , 2$. The $8\times 8$ matrix $\Omega_{AB}$ matrix is then block diagonal,
\begin{equation}
(\Omega_{AB}) = \begin{pmatrix} \Omega_{\tta\ttb} & 0 \\ 0 & \varepsilon_{\alpha\beta} \end{pmatrix} \,.
\end{equation}
The corresponding splitting of the fields is then
\begin{eqnarray}
\Theta^{a}= \theta^{a}\, , \quad \quad \Theta^{\alpha} = \theta^{\alpha} \, ,
\end{eqnarray}
and
\begin{align}
  W^{\tta \ttb}_{ij}&= w^{\tta \ttb}_{ij} +\frac{1}{2\sqrt{6}} \Omega^{\tta \ttb} w_{ij}\, , & W^{\tta \alpha}_{ij} &=\frac{1}{\sqrt{2}} w^{\tta \alpha}_{ij} \, ,&  W^{\alpha \beta} &= -\frac{3}{2\sqrt{6}} w_{ij} \varepsilon^{\alpha \beta}  \, , \\
  V^{\tta \ttb}_{i}&= v^{\tta \ttb}_{i} +\frac{1}{2\sqrt{6}} \Omega^{\tta \ttb} v_{i} \, , & V^{\tta \alpha}_{i} &=\frac{1}{\sqrt{2}} v^{\tta \alpha}_{i} \, , & V^{\alpha \beta} &= -\frac{3}{2\sqrt{6}} v_{i} \varepsilon^{\alpha \beta}  \, , \\
 \Psi^{\tta \ttb \ttc} &= \psi^{\tta \ttb \ttc}+\frac{1}{2} \Omega^{[\tta \ttb}\psi^{\ttc]}\, ,& \Psi^{\tta \ttb \alpha} &= \frac{1}{\sqrt{3}} \psi^{\tta \ttb \alpha} \, , & \Psi^{\tta \alpha \beta} &= -\frac{1}{3} \psi^{\tta} \varepsilon^{\alpha \beta}\, ,\\
  \Phi^{\tta \ttb \ttc \ttd} &= -\sqrt{\frac{3}{2}} \Omega^{[\tta \ttb}\phi^{\ttc \ttd]} \, ,& \Phi^{\tta \ttb \ttc \alpha} &= \frac{1}{2}\phi^{\tta \ttb \ttc \alpha} \, , & \Phi^{\tta \ttb \alpha \beta} &= \frac{1}{2\sqrt{6}} \phi^{\tta \ttb} \varepsilon^{\alpha \beta} \, ,\\
   \Pi^{\tta \ttb \ttc \ttd} &= -\sqrt{\frac{3}{2}} \Omega^{[\tta \ttb}\pi^{\ttc \ttd]} \, , & \Pi^{\tta \ttb \ttc \alpha} &= \frac{1}{2}\pi^{\tta \ttb \ttc \alpha} \, , & \Pi^{\tta \ttb \alpha \beta} &= \frac{1}{2\sqrt{6}} \pi^{\tta \ttb} \varepsilon^{\alpha \beta} \, .
\end{align}
These coefficients are determined uniquely (up to signs that can be absorbed by field redefinitions) from the following two conditions: 1) the compatibility of $\mathfrak{usp}(8)$ and $\mathfrak{usp}(6)\oplus \mathfrak{usp}(2)$ irreducibility conditions, and 2) the normalization of the action.
For example, for the scalar fields, those two conditions are
\begin{equation}
\Omega_{AB} \Phi^{ABCD} = 0 \quad \Leftrightarrow \quad \Omega_{\tta\ttb} \phi^{\tta\ttb} = 0\,, \; \Omega_{\tta\ttb}\phi^{\tta\ttb\ttc \alpha} = 0
\end{equation}
and
\begin{equation}
\Phi^*_{ABCD} \Phi^{ABCD} = \phi^*_{\tta\ttb\ttc \alpha} \phi^{\tta\ttb\ttc \alpha} + \phi^*_{\tta\ttb} \phi^{\tta\ttb}\, .
\end{equation}
Reality conditions are also compatible. Using these formulas, the reduction of the $\cN = (3,1)$ theory gives exactly the action of maximal supergravity in five dimensions presented in subsection \ref{sec:5daction}.

\paragraph{Comparison of supersymmetry transformations.} It is also straightforward to reduce the supersymmetry transformations of section \ref{sec:susy} to five dimensions. On the other hand, we can use this splitting of indices in the supersymmetry transformations of section \ref{sec:SUSY40}. Comparing the two enables us to fix the $\beta_i$ constants of the $\cN = (3,1)$ theory as
\begin{align}
\beta_1    & = \frac{1}{\sqrt{3}} \alpha_1 = -6 \,\alpha_2 = 2\sqrt{6} \,\alpha_3 \, ,            &\beta_6    & = \frac{1}{\sqrt{6}}\alpha_3 \, ,                                                    \\
\beta_2    & = -\frac{\sqrt{2}}{3\sqrt{3}} \alpha_1 = 2\sqrt{2}  \alpha_2 \, ,                    &\beta_7    & = \frac{1}{\sqrt{6}} \alpha_3 = \frac{1}{2\sqrt{2}} \alpha_4 \, ,                    \\
\beta_3    & = -\frac{\alpha_2}{2}= \frac{1}{\sqrt{6}} \alpha_3  \, ,                             &\beta_8    & = \frac{1}{\sqrt{2}} \alpha_3 = \frac{\sqrt{3}}{2\sqrt{2}} \alpha_4 \, ,             \\
\beta_4    & = -\frac{\alpha_2}{2}= \frac{1}{\sqrt{6}} \alpha_3=\frac{1}{2\sqrt{2}} \alpha_4 \, , &\beta_9    & = -\frac{\sqrt{3}}{2} \alpha_4 \, ,                                                  \\
\beta_5    & = -\frac{1}{\sqrt{2}} \alpha_2 \, ,                                                  &\beta_{10} & = -\frac{\alpha_4}{2} \, ,                                                             
\end{align}
in terms of the $\alpha_i$ constants of the $\cN = (4,0)$ theory. This is compatible with relations \eqref{eq:relalphas}, and gives the relations \eqref{eq:relbetas} announced in section \ref{sec:susy}.

\providecommand{\href}[2]{#2}\begingroup\raggedright\endgroup

\end{document}